# Symmetry and Degeneracy of Phonon Modes for Periodic Structures with Glide Symmetry

Pu Zhang[1]

*Department of Mechanical Engineering, State University of New York at Binghamton, Binghamton, NY 13902, United States*

**Abstract:** A large class of phononic crystals and mechanical metamaterials exhibit glide symmetry that dictates their functionality or exceptional performance. The glide symmetry gives rise to a number of intriguing phenomena like sticking-bands and degeneracy in the phononic band structures. Fully understanding of these phenomena demands analysis of the phonon modes' symmetry property, which is, however, a challenging task since it involves nonsymmorphic space group analysis and special treatment of the Brillouin zone boundary. Therefore, this work introduces a systematic group-theoretical procedure determining the symmetry of phonon modes for periodic structures with glide symmetry. By taking the $p4g$ group as an example, the symmetry of phonon modes is discussed by deriving the small representations for high symmetry **k**-points, and different types of degeneracies are elucidated. This work provides insight into the role of glide symmetry on phononic band structures and guides the symmetry analysis of periodic structures of other types.

## 1 Introduction

Recent progress on the design of structured materials mainly depends on special geometry to achieve desirable or exceptional functionality. A variety of novel structures and mechanisms have been summarized in some recent review articles (Bertoldi et al., 2017; Hussein et al., 2014; Lakes, 2017; Lee et al., 2012; Saxena et al., 2016; Yu et al., 2018; Zadpoor, 2016) and books (Laude, 2015; Phani and Hussein, 2017) regarding phonon crystals, mechanical/acoustic metamaterials, et al. In the literature, plenty of designs have involved structures with glide symmetry, either intentionally or not, that plays an essential role on their exceptional behaviors. Some of the representative structures with glide symmetry are illustrated in Figure 1. The first example is the hinged rotating squares (Figure 1a) with auxetic behavior (Grima et al., 2005; Grima and Evans, 2000). Normally the square patches are made of rigid materials. This structure has inspired a few similar designs, e.g. the fractal cut structures with supreme stretchability (Cho et al., 2014; Dudek et al., 2017; Tang et al., 2015) or controllable dynamic behaviors (Javid et al., 2016). The example shown in Figure 1b is obtained from the buckling pattern of an elastomer sheet with a square array of pores under biaxial compression (Bertoldi et al., 2008; Mullin et al., 2007). The glide symmetry of the porous sheet stems from symmetry breaking of the structure during the buckling process. This design has been followed by a few fundamental and applied studies on wave control (Bertoldi and Boyce, 2008) and auxetic/reconfigurable materials (Bertoldi et al., 2010; Shim et al., 2013). Recently, a few works published on programmable mechanical metamaterials also exhibit similar geometry as the porous sheet (Coulais, 2016; Florijn et al., 2014). The example in Figure 1c shows a sinusoidal lattice structure obtained from the buckling mode of a square lattice structure (Haghpanah et al., 2014; Ohno et al., 2002). This sinusoidal structure has been studied by a few researchers regarding the auxetic behavior (Körner and Liebold-Ribeiro, 2015), wave propagation characteristics (Chen et al., 2017b; Trainiti et al., 2016), or synergy behavior (Chen et al., 2017a). In addition, a similar structure of this type was obtained after swelling instability of a soft lattice structure (Liu et al., 2016). Figure 1d shows an anti-tetrachiral structure proposed in (Grima et al., 2008) but was originally discovered from topology optimization (Sigmund et al., 1998). The mechanical behavior of this structure is not thoroughly explored yet but some works on the static properties have been reported (Alderson et al., 2010; Clausen et al., 2015). While all of the aforementioned four structures are in the plane group *p*4*g*, structures belonging to other groups with glide symmetry can also be found in the literature. For instance, Figure 1e shows a sinusoidal beam (Maurin and Spadoni, 2014; Trainiti et al., 2015) that is in the Frieze group *p*2*mg* (Zhang and Parnell, 2017a). A similar zig-zag structure proposed in (Nanda and Karami, 2018) also belongs to this group. At last, Figure 1f illustrates a structure with glide symmetry that is obtained from the buckling mode of a honeycomb structure (Haghpanah et al., 2014; Ohno et al., 2002). This structure belongs to the plane group *pgg* and is named as 'anti-rolls' (Saiki et al., 2005).

---

[1] Email: pzhang@binghamton.edu

Many of the aforementioned works are on the wave propagation behavior with a focus on the formation and/or tunability of phononic band gaps (Bertoldi and Boyce, 2008; Chen et al., 2017a, 2017b; Javid et al., 2016; Lustig and Shmuel, 2018; Trainiti et al., 2016, 2015; Zhang and Parnell, 2017a, 2017b). However, very little attention has been paid on the effect of glide symmetry on the characteristics of band structures. A unique feature of the band structure for structures with glide symmetry is the sticking-bands effect at the boundary of the first Brillouin zone (FBZ). This phenomenon has long been known for crystals with glide symmetry (Dresselhaus et al., 2008) and can also be found for the phononic band structures of continuum structures (Bertoldi and Boyce, 2008; Chen et al., 2017a; Trainiti et al., 2015; Zhang and Parnell, 2017). The sticking-bands effect is directly induced by the degeneracy of phonon modes, which could be deciphered by conducting group-theoretical analysis of the symmetry of phonon modes. In addition, analyzing the symmetry of phonon modes also offers a reliable way to sort phonon bands by identifying the band crossing and anticrossing (or avoided crossing) correctly, which is a challenging process even when the glide symmetry vanishes (Lu and Srivastava, 2018). Another potential impact of studying the symmetry of phonon modes is to design novel phononic crystals with broad or controllable band gaps by taking the advantage of glide symmetry. Although this approach is still emerging, some research has shed light on the promising future of nonsymmorphic phononic crystals. For example, it was reported that glide symmetry helps create gigantic complete band gaps by coalescing bands near the FBZ edge together and reducing the curvatures of some bands (Koh, 2011). Further discussion on the design principles employing glide symmetry can be found in a recent book (Gan, 2017). In order to transform the group theory from an analysis method to a design tool for phononic crystals, in-depth understanding of the symmetry of phonon modes and the degeneracies must be carried out first, which is a major motivation of this work.

The key issue in determining the symmetry of phonon modes is to find the corresponding small representations for each band by using the group theory. It is relatively easy to find the small representations for symmorphic structures, i.e. without glide/screw symmetries, by following the procedure introduced by (Sakoda, 2005). Specific examples on analyzing the symmetry of photon modes can be found in (Alagappan et al., 2008; Hergert and Däne, 2003) for symmorphic photonic crystals. In contrast, it is very challenging to conduct symmetry analysis of phonon modes for nonsymmorphic structures once the glide/screw symmetry exists. For this reason, systematic studies on nonsymmorphic structures are very rare in the literature, although detailed analysis of one dimensional (1D) structures can be found (Mock et al., 2010). Therefore, this work aims at filling this gap by introducing a systematic procedure determining the symmetry of phonon modes for structures with glide symmetry. The group theoretical methods used in this work can be found in some textbooks (Altmann, 2002; Bradley and Cracknell, 2010; Inui et al., 1990; Jacobs, 2005). We focus on structures in the *p*4*g* group since they are the most common and useful ones in the literature as shown in Figure 1. In addition, only the sinusoidal lattice structure in Figure 1c is studied in this work since it is easy to illustrate its phonon modes. Other *p*4*g* structures exhibit the same symmetry behavior as the sinusoidal lattice so all the small representations are identical.

## 2  Lattice Structure in p4g Group

### 2.1  Symmetry of the Structure

In order to study the symmetry characteristics of lattice structures with glide symmetry in two dimension (2D), we consider a lattice structure composed of interconnected sinusoidal beams illustrated in Figure 2, which belongs to the square lattice class. The primitive lattice vectors are represented by $\mathbf{a}_1$ and $\mathbf{a}_2$, respectively, with lattice constants as *a* along both directions. Note that some researchers (Bertoldi and Boyce, 2008; Chen et al., 2017a, 2017b; Javid et al., 2016; Trainiti et al., 2016) adopt a non-primitive unit cell that is shown in Figure 2 to calculate the band structures. These two approaches certainly generate identical band gaps. Nevertheless, the phononic bands obtained from the non-primitive unit cells are usually more complicated than those obtained from the primitive unit cells, which is because that there are two lattice points in one non-primitive unit cell so that the corresponding FBZ will be folded with more bands crossing each other. Therefore, the primitive unit cell is recommended when dealing with the symmetry behaviors of phonon modes.

The symmetry of the lattice structure in the real space is shown in Figure 3a. This structure belongs to the nonsymmorphic plane group *p4g* and the point group 4*mm* (Aroyo, 2016; Bradley and Cracknell, 2010). It is known that the point group 4*mm* has an order of 8, which indicates that there are a total of 8 distinct types of symmetry operations. As shown in Figure 3a, the center of the unit cell is a 4-fold rotation axis with four symmetry operations, namely, identity operation $\{E|\mathbf{0}\}$, 90° rotation $\{C_4^+|\mathbf{0}\}$, 180° rotation $\{C_2|\mathbf{0}\}$, and –90° rotation $\{C_4^-|\mathbf{0}\}$. Besides, there are four types of glide-reflection lines along the axial directions ($\{\sigma_x|\boldsymbol{\tau}\}, \{\sigma_y|\boldsymbol{\tau}\}$) and diagonal directions ($\{\sigma_d|\boldsymbol{\tau}\}, \{\sigma_d'|\boldsymbol{\tau}\}$), respectively, with $\boldsymbol{\tau}=(\mathbf{a}_1+\mathbf{a}_2)/2$ representing the sub-lattice glide translation vector. Note that we adopt the Seitz operator (Inui et al., 1990) to denote a space group operation, as

$$\{\mathbf{R}|\mathbf{t}\}\mathbf{x} = \mathbf{R}\mathbf{x} + \mathbf{t}, \tag{1}$$

where $\mathbf{R}$ and $\mathbf{t}$ represent the orthogonal and translation transformations, respectively, and $\mathbf{x}$ is a spatial point in the structure. Conversely, an inverse transformation of Eq. (1) is defined as

$$\{\mathbf{R}|\mathbf{t}\}^{-1}\mathbf{x} = \mathbf{R}^{-1}(\mathbf{x}-\mathbf{t}). \tag{2}$$

Note that Eq. (1) and Eq. (2) will degenerate to the corresponding point group operations when the translation vector $\mathbf{t}$ vanishes. Applying the Seitz operator to a scalar field $f(\mathbf{x})$ and a vector field $\mathbf{f}(\mathbf{x})$ will result in the following transformation by definition,

$$\begin{aligned}\{\mathbf{R}|\mathbf{t}\}f(\mathbf{x}) &= f(\{\mathbf{R}|\mathbf{t}\}^{-1}\mathbf{x}), \\ \{\mathbf{R}|\mathbf{t}\}\mathbf{f}(\mathbf{x}) &= \mathbf{R}\mathbf{f}(\{\mathbf{R}|\mathbf{t}\}^{-1}\mathbf{x}).\end{aligned} \tag{3}$$

Equation (3) indicates that a space group transformation to a vector field $\mathbf{f}(\mathbf{x})$ will change both the argument and the function itself, whereas for a scalar field $f(\mathbf{x})$ only the argument needs to be transformed.

The plane group $\mathcal{G}$ of this lattice structure consists all of the plane group transformations represented by Eq. (1) that leave the structure invariant. Mathematically, this plane group is expressed as

$$\begin{aligned}\mathcal{G} = &\mathcal{T}\{E|\mathbf{0}\} + \mathcal{T}\{C_4^+|\mathbf{0}\} + \mathcal{T}\{C_2|\mathbf{0}\} + \mathcal{T}\{C_4^-|\mathbf{0}\} \\ &+ \mathcal{T}\{\sigma_x|\boldsymbol{\tau}\} + \mathcal{T}\{\sigma_y|\boldsymbol{\tau}\} + \mathcal{T}\{\sigma_d|\boldsymbol{\tau}\} + \mathcal{T}\{\sigma_d'|\boldsymbol{\tau}\},\end{aligned} \tag{4}$$

where $\mathcal{T}$ is the translation group consisting all lattice translation operations $\mathbf{a}_n = n_1\mathbf{a}_1 + n_2\mathbf{a}_2$ with integer $n_1$ and $n_2$. Each of the 8 cosets in Eq. (4) denotes one type of symmetry for this lattice structure. Note that the 8 operations ($\{E|\mathbf{0}\}$, $\{C_4^+|\mathbf{0}\}$, ..., $\{\sigma_d'|\boldsymbol{\tau}\}$) do not form the corresponding point group 4*mm* for the plane group *p*4*g* due to the glide symmetry. Rather than that, only the factor group $\mathcal{G}/\mathcal{T}$ is isomorphic to 4*mm*. This is exactly why plane/space group analysis is necessary for periodic structures in the nonsymmorphic groups.

## 2.2 Symmetry in the k-Space

It is usually quite convenient to study the phonons in the reciprocal space (**k**-space), where **k** indicates the wave vector. As shown in Figure 3b, the reciprocal lattice of the *p*4*g* structure is in the square class with reciprocal lattice vectors $\mathbf{b}_1$ and $\mathbf{b}_2$ defined as

$$\mathbf{b}_i = \frac{2\pi}{a^2}\mathbf{a}_i \quad (i=1,2). \tag{5}$$

Accordingly, the wave vector **k** is defined as $\mathbf{k} = k_x\mathbf{b}_1 + k_y\mathbf{b}_2$. The **k**-space is periodic so normally we only need to consider the primitive unit cell, which is known as the FBZ. In addition, different from the real space that may be nonsymmorphic, the **k**-space is always symmorphic and therefore only the point group symmetry needs to be considered in the FBZ.

The point group of a **k**-point (Altmann, 2002; Inui et al., 1990) is defined as the symmetry operations that transform the **k**-point into itself or an equivalent **k**-point $\mathbf{k}+\mathbf{b}_n$, where $\mathbf{b}_n = n_1\mathbf{b}_1 + n_2\mathbf{b}_2$ represents an arbitrary reciprocal lattice vector with integers $n_1$ and $n_2$. More specifically, the interior **k**-points of the FBZ will always be transformed into themselves by the point group operations, while the **k**-points on the boundaries of the FBZ may be transformed into their equivalent **k**-points (Altmann, 2002; Inui et al., 1990). Consider the $\Gamma$ point $\mathbf{k} = (0,0)$ in Figure 3b as an example, all of the 8 operations ($E$, $C_4^+$, $C_2$, $C_4^-$, $\sigma_x$, $\sigma_y$, $\sigma_d$, $\sigma_d'$) in the 4*mm* point group will leave the $\Gamma$ point invariant so that the point group of $\Gamma$ is 4*mm*, the same as that of the plane group *p*4*g*. In contrast, only two of the operations ($E$, $\sigma_y$) will leave the $\Delta$ point invariant, forming a point group *m* which is actually a subgroup of 4*mm*. For the points on the FBZ boundaries, such as *X*, *Y* and *M*, the corresponding symmetry operations may transform them into their equivalent **k**-points, which can be verified easily. The point group of each high symmetry **k**-point and the corresponding symmetry operations are tabulated in Table 1, where the location is denoted as $(m_1\ m_2) = m_1\mathbf{b}_1 + m_2\mathbf{b}_2$. Note that the point group of $\Gamma$ exhibits the highest symmetry so it is defined as the point group of the corresponding plane/space group. On the other hand, the point groups of all other high symmetry points are merely subgroups of that for the $\Gamma$ point. Due to the symmetry of the FBZ, we can further reduce it to 1/8 of the original size, which is called the irreducible FBZ and illustrated in Figure 3b. Normally the band structures and phonon modes are investigated just for the irreducible FBZ.

Even though the point group of **k** is defined in the **k**-space, the group of **k** (Altmann, 2002; Inui et al., 1990) is indeed defined in the real space, which characterizes how the plane group operations affect the **k** point. By definition, the group of **k**, denoted as $\mathcal{G}^{\mathbf{k}}$, consists all the plane group operations in $\mathcal{G}$ that transform the **k**-point to itself or an equivalent **k**-point. Hence, $\mathcal{G}^{\mathbf{k}}$ is a subgroup of the plane group $\mathcal{G}$. Mathematically, one can simply choose the cosets in Eq. (4) that correspond to the point group of **k** in Table 1. For example, it is easy to verify that $\mathcal{G}^{\mathbf{k}} = \mathcal{G}$ for the points $\Gamma$ and *M*. While for the *X* point, the group of **k** is

$$\mathcal{G}^{\mathbf{k}}(X) = \mathcal{T}\{E|\mathbf{0}\} + \mathcal{T}\{C_2|\mathbf{0}\} + \mathcal{T}\{\sigma_x|\boldsymbol{\tau}\} + \mathcal{T}\{\sigma_y|\boldsymbol{\tau}\} . \tag{6}$$

Moreover, the group of **k** for the remaining three high symmetry points can be defined in a similar manner as

$$\begin{aligned} \mathcal{G}^{\mathbf{k}}(\Delta) &= \mathcal{T}\{E|\mathbf{0}\} + \mathcal{T}\{\sigma_y|\boldsymbol{\tau}\}, \\ \mathcal{G}^{\mathbf{k}}(\Sigma) &= \mathcal{T}\{E|\mathbf{0}\} + \mathcal{T}\{\sigma'_d|\boldsymbol{\tau}\}, \\ \mathcal{G}^{\mathbf{k}}(Y) &= \mathcal{T}\{E|\mathbf{0}\} + \mathcal{T}\{\sigma_x|\boldsymbol{\tau}\}. \end{aligned} \tag{7}$$

Note that the group of **k** is needed for nonsymmorphic groups to study the feature of phonon modes, while the point group of **k** is usually enough for symmorphic groups (Inui et al., 1990).

Note that the symmetry of phonon modes only needs to be analyzed for high symmetry **k** points/lines. For general **k** points with translation symmetry only, the group of **k** is $\mathcal{G}^{\mathbf{k}} = \mathcal{T}\{E|\mathbf{0}\}$ so the phonon mode only satisfies translation symmetry, i.e. the Bloch condition (Bradley and Cracknell, 2010). However, it is suggested to consider all the **k** points when determining the band gaps of a phononic crystal since the band gap extremum does not necessarily locate at high symmetry **k** points/lines (Kerszberg and Suryanarayana, 2015; Maurin et al., 2018; Pratapa et al., 2018).

## 2.3   Phonon Mode Calculation

The phonon modes of the lattice structure in Figure 2 is calculated by using finite element analysis. Each ligament is considered as a curved Timoshenko beam with three degree-of-freedoms, i.e. horizontal displacement $u_x$, vertical displacement $u_y$, and in-plane rotation angle $\theta$. If we denote the wave field as $\psi = (u_x \ u_y \ \theta)$, the Bloch wave solution is

$$\psi(\mathbf{x},t) = \psi_{\mathbf{k}}(\mathbf{x})e^{-i\omega t} , \tag{8}$$

where $i = \sqrt{-1}$, $t$ is the time, and $\psi_{\mathbf{k}}(\mathbf{x})$ is the phonon/Bloch mode corresponding to the wave vector **k** and angular frequency $\omega$. The phonon mode $\psi_{\mathbf{k}}(\mathbf{x})$ is a complex function in general, although only the real or imaginary part is depicted in some literature. Further, the phonon mode $\psi_\mathbf{k}(\mathbf{x})$ satisfies the Bloch condition (Bradley and Cracknell, 2010), as

$$\psi_\mathbf{k}(\mathbf{x}) = \tilde{\psi}_\mathbf{k}(\mathbf{x})e^{i\mathbf{k}\mathbf{x}}, \tag{9}$$

where $\tilde{\psi}_\mathbf{k} = (\tilde{u}_x \ \tilde{u}_y \ \tilde{\theta})_\mathbf{k}$ is a spatial periodic function and invariant under an arbitrary lattice translation. Mathematically, this periodic boundary condition of $\tilde{\psi}_\mathbf{k}$ is defined as

$$\tilde{\psi}_\mathbf{k}(\mathbf{x}) = \tilde{\psi}_\mathbf{k}(\mathbf{x} + \mathbf{a}_n), \tag{10}$$

where $\mathbf{a}_n = n_1 \mathbf{a}_1 + n_2 \mathbf{a}_2$ indicates an arbitrary lattice translation with $n_i \ (i=1,2)$ as integers.

The method proposed in (Åberg and Gudmundson, 1997) is adopted to implement the finite element analysis. Following this method, two identical models are established for the unit cell in Figure 3a, one for the real solution $\psi_\mathbf{k}^{Re}(\mathbf{x})$ and the other for the imaginary solution $\psi_\mathbf{k}^{Im}(\mathbf{x})$ of the complex eigenmodes $\psi_\mathbf{k}(\mathbf{x}) = \psi_\mathbf{k}^{Re}(\mathbf{x}) + i\psi_\mathbf{k}^{Im}(\mathbf{x})$. It can be proved from Eq. (9) and Eq. (10) that the real eigenmode $\psi_\mathbf{k}^{Re}(\mathbf{x})$ and imaginary eigenmode $\psi_\mathbf{k}^{Im}(\mathbf{x})$ satisfy the following Bloch condition, as

$$\begin{aligned}\psi_\mathbf{k}^{Re}(\mathbf{x} - \mathbf{a}_n) &= \cos(\mathbf{k}\mathbf{a}_n)\psi_\mathbf{k}^{Re}(\mathbf{x}) + \sin(\mathbf{k}\mathbf{a}_n)\psi_\mathbf{k}^{Im}(\mathbf{x}), \\ \psi_\mathbf{k}^{Im}(\mathbf{x} - \mathbf{a}_n) &= -\sin(\mathbf{k}\mathbf{a}_n)\psi_\mathbf{k}^{Re}(\mathbf{x}) + \cos(\mathbf{k}\mathbf{a}_n)\psi_\mathbf{k}^{Im}(\mathbf{x}).\end{aligned} \tag{11}$$

Consequently, the Bloch condition (11) can be implemented easily for all the boundary and vertex nodes in Figure 3a. Finally, the band structures are obtained by solving the eigenfrequencies $\omega_\mathbf{k}$ for each $\mathbf{k}$-point and the corresponding eigenmodes can be extracted as well to compute the phonon modes.

The computational models are implemented in the commercial finite element software ABAQUS 2017 (Simulia, 2017) to conduct the simulation. The Timoshenko beam element B21 is used with a total of 104 elements divided for the whole model. The unit cell length is taken as $a = 1/\sqrt{2}$ m for the square lattice. Each of the four ligaments is in fact a half period of sinusoidal curve with an amplitude of 0.15 m. The cross section of the beam is square with a thickness of $h = 0.005$ m. The material chosen here is stainless steel with elastic modulus as $E = 200$ GPa, Poisson's ratio as $\nu = 0.3$, and density as $\rho = 7900 \text{ kg/m}^3$. A dimensionless frequency $\bar{\omega}$ is introduced for all the results shown in this work, as

$$\bar{\omega} = \omega a \sqrt{\frac{\rho A}{EI}}, \tag{12}$$

where $A = h^2$ and $I = h^4/12$ are the area and moment of inertia for the cross section of the beam. Note that we constrain the motion of the lattice structure in 2D.

# 3 Phonon Mode Symmetry for p4g Lattice

## 3.1 General Theory

The key issue in uncovering the symmetry of phonon modes is to find the small (or little) representation for $\mathcal{G}^\mathbf{k}$, the group of $\mathbf{k}$, since the Bloch wave functions serve as bases for the representation of $\mathcal{G}^\mathbf{k}$ (Inui et al., 1990). Normally only the high symmetry $\mathbf{k}$-points need to be considered, which include interior points ($\Gamma$, $\Delta$, $\Sigma$) and boundary points ($X$, $Y$, $M$) for nonsymmorphic structures in the $p4g$ group. Since only the displacements $u_x$ and $u_y$ are considered for the symmetry analysis, we define a modified Bloch wave function $\phi_\mathbf{k} = (u_x\ u_y)_\mathbf{k}$. Based on the group theory (Inui et al., 1990), the small representation of $\mathcal{G}^\mathbf{k}$ is

$$\{\mathbf{R}\,|\,\mathbf{t}\}\phi_{\mathbf{k}i} = \sum_{j=1}^{d} \phi_{\mathbf{k}j} D_{ji}^\mathbf{k} \ , \tag{13}$$

where $\{\mathbf{R}\,|\,\mathbf{t}\}$ is a plane group transformation, $\mathbf{R}$ belongs to the point group of $\mathbf{k}$ listed in Table 1 for a $\mathbf{k}$-point, $\mathbf{t}$ is the translation vector, $\phi_{\mathbf{k}i}$ is the phonon mode to be studied, and $D_{ji}^\mathbf{k}$ is a representation matrix for $\{\mathbf{R}\,|\,\mathbf{t}\}$, and $d \geq 1$ is the dimension of the representation matrix. Some special remarks are noted here for the specific form of $D_{ji}^\mathbf{k}$ (Bradley and Cracknell, 2010; Inui et al., 1990).

1) For a 1D representation ($d = 1$), Eq. (13) is simplified as $\{\mathbf{R}\,|\,\mathbf{t}\}\phi_\mathbf{k} = D^\mathbf{k}\phi_\mathbf{k}$ with $|D^\mathbf{k}| = 1$. In this case, we only need to consider the symmetry of a phonon mode $\phi_\mathbf{k}$ with itself. There is no frequency degeneracy induced by the symmetry, although accidental degeneracy may still occur.

2) For higher dimensional representation ($d > 1$), $\phi_{\mathbf{k}i}$ ($i = 1, 2, \cdots, d$) are a series of linearly independent Bloch modes corresponding to the same frequency. Hence the frequency degeneracy is guaranteed by the symmetry in this case. In addition, these Bloch modes interact and couple with each other under symmetry transformations.

3) For lattice translation operation $\{E\,|\,\mathbf{a}_n\}$, the representation matrix $D_{ji}^\mathbf{k}$ obeys the following form as

$$D_{ji}^\mathbf{k}(\{E\,|\,\mathbf{a}_n\}) = \exp(-i\mathbf{k}\mathbf{a}_n)\mathbf{I}_d \ , \tag{14}$$

where $\mathbf{I}_d$ is the unit matrix with dimension $d$. Equation (14) arises from the representation for the translation group $\mathcal{T}$. The relation in (14) is useful to rule out spurious representations when the Herring's method (Inui et al., 1990; Jacobs, 2005) is used to derive small representations for boundary $\mathbf{k}$-points, which will be introduced in Section 3.3.

The major task to study the symmetry of phonon modes is to find the small representation matrices $D_{ji}^{\mathbf{k}}$ in Eq. (13) for each high symmetry **k**-point. However, the analyses to the interior and boundary points in the FBZ are quite different, which should be considered separately.

### 3.2 Interior k-Points

For any **k**-points inside the FBZ, the small representation matrix $D_{ji}^{\mathbf{k}}$ can be expressed as

$$D_{ji}^{\mathbf{k}}(\{\mathbf{R}\mid\mathbf{t}\}) = \exp(-i\mathbf{k}\mathbf{t})D_{ji}(\mathbf{R}) \ , \tag{15}$$

where $D_{ji}(\mathbf{R})$ is the irreducible representation matrix for the point group of **k**, which can be found in the table of point groups (Altmann and Herzig, 1994). Hence by considering Eq. (15), the small representation Eq. (13) can be further derived as

$$\{\mathbf{R}\mid\mathbf{t}\}\phi_{\mathbf{k}i} = \exp(-i\mathbf{k}\mathbf{t})\sum_{j=1}^{d}\phi_{\mathbf{k}j}D_{ji}(\mathbf{R}) \ , \tag{16}$$

where $\exp(-i\mathbf{k}\mathbf{t})$ is the phonon mode phase factor induced by the translation $\mathbf{t}$, which will vanish for symmorphic groups. Based on Eq. (16), the phonon modes for the three interior points ($\Gamma$, $\Delta$, $\Sigma$) are investigated below.

The $\Gamma$ point locates at the origin of the FBZ with $\mathbf{k} = \mathbf{0}$ so that the phase factor in Eq. (16) vanishes. Thus, one just needs to use the representation table of the point group $4mm$, which is shown in Table 2. Among the five small representations $\Gamma_i$ ($i = 1,2,\cdots 5$), the first four are all 1D while the representation $\Gamma_5$ is 2D. This indicates that the phonon modes may have double degeneracies at the $\Gamma$ point if the representation is $\Gamma_5$ for any bands. Besides the characters of $D_{ji}^{\mathbf{k}}$, we also present the representation matrices in Table 2 by defining the following four orthogonal matrices, as

$$\mathbf{I} = \begin{bmatrix} 1 & 0 \\ 0 & 1 \end{bmatrix}, \quad \mathbf{A} = \begin{bmatrix} 1 & 0 \\ 0 & -1 \end{bmatrix}, \quad \mathbf{B} = \begin{bmatrix} 0 & 1 \\ 1 & 0 \end{bmatrix}, \quad \mathbf{C} = \begin{bmatrix} 0 & 1 \\ -1 & 0 \end{bmatrix}. \tag{17}$$

The $\Delta$ point is defined as $\mathbf{k} = (\alpha,0)$ with $0 < \alpha < 1/2$. There are only two symmetry operations $\{E\mid\mathbf{0}\}$ and $\{\sigma_y\mid\boldsymbol{\tau}\}$ for this **k**-point. For the glide symmetry $\{\sigma_y\mid\boldsymbol{\tau}\}$, the phase factor in Eq. (16) is calculated as $\exp(-i\mathbf{k}\boldsymbol{\tau}) = \exp(-i\alpha\mathbf{b}_1 \cdot \tfrac{1}{2}\mathbf{a}_1) = \exp(-\alpha\pi i)$. Thus the small representations can be obtained by modifying the representation table of the point group $m$, as shown in Table 3. Both $\Delta_1$ and $\Delta_2$ are 1D representations so there is no frequency degeneracy induced by the symmetry.

The analysis of the $\Sigma$ point is similar to the $\Delta$ point. For the **k**-points with $\mathbf{k}=(\alpha,\alpha)$, only two symmetry operations $\{E|\mathbf{0}\}$ and $\{\sigma'_d|\boldsymbol{\tau}\}$ exist. The phase factor for the latter is $\exp(-i\mathbf{k}\boldsymbol{\tau})=\exp(-2\alpha\pi i)$. The small representations can be obtained in Table 4. Both $\Sigma_1$ and $\Sigma_2$ are 1D representations.

So far, we have presented the small representations $D^{\mathbf{k}}_{ji}$ of the interior points $\Gamma$, $\Delta$, and $\Sigma$ in Table 2, Table 3, and Table 4, respectively. Consequently, the symmetry of the phonon modes can be obtained by using Eq. (13). Note that the condition (14) is satisfied automatically for all the representations of interior **k**-points due to the existence of the phase factor $\exp(-i\mathbf{k}\mathbf{t})$ in Eq. (16).

### 3.3 Boundary k-Points

For nonsymmorphic structures, the small representation for **k**-points at the boundary of the FBZ is usually complicated. There are two types of boundary **k**-points (Inui et al., 1990), which should be treated separately, as

*Type I: Boundary **k**-points with symmorphic $\mathcal{G}^{\mathbf{k}}$, i.e. without involving any essential glides/screws.*
*Type II: Boundary **k**-points with nonsymmorphic $\mathcal{G}^{\mathbf{k}}$.*

For boundary **k**-points of type I, Eq. (16) is still valid so the symmetry of phonon modes can be analyzed similarly to the interior **k**-points. However, more complicated techniques are required for boundary **k**-points of type II, e.g. the Herring's method or the theory of ray representations (Inui et al., 1990).

It is inferred from Eqs. (6) and (7) that the $\mathcal{G}^{\mathbf{k}}$ of all boundary **k**-points ($X, Y, M$) for the $p4g$ group involves certain glide symmetry operation. Therefore, all the boundary **k-**points belong to the type II classified above. The Herring's method (Inui et al., 1990; Jacobs, 2005) is adopted in this work since it is convenient to use. The first step of implementing Herring's method is to find a set of lattice translations $\mathcal{T}^{\mathbf{k}}$ that satisfies the following relation, as

$$\exp(-i\mathbf{k}\mathbf{a}_n)=1, \tag{18}$$

or equivalently,

$$\mathbf{k}\mathbf{a}_n=\mathbf{k}(n_1\mathbf{a}_1+n_2\mathbf{a}_2)=2n\pi, \tag{19}$$

where $n_1$, $n_2$, and $n$ are arbitrary integers. In fact, $\mathcal{T}^{\mathbf{k}}$ defined by (18) or (19) is a subgroup of $\mathcal{T}$. Once the translation group $\mathcal{T}^{\mathbf{k}}$ is obtained, the second step is to derive the factor group $\mathcal{G}^{\mathbf{k}}/\mathcal{T}^{\mathbf{k}}$, which can be obtained by decomposing $\mathcal{G}^{\mathbf{k}}$ with the cosets of $\mathcal{T}^{\mathbf{k}}$. Finally, the small representation $D^{\mathbf{k}}_{ji}$ in Eq. (13) can be obtained from the irreducible representations of the factor group $\mathcal{G}^{\mathbf{k}}/\mathcal{T}^{\mathbf{k}}$ for the boundary **k**-points of type II. The Herring's method is very powerful in determining the small representations for the boundary

**k**-points of nonsymmorphic groups. However, a drawback is that not all of the irreducible representations of $\mathcal{G}^k/\mathcal{T}^k$ satisfy the condition (14). These irreducible representations are spurious solutions and not permissible to serve as $D_{ji}^k$ in Eq. (13).

For the $X$ point with $\mathbf{k} = (\frac{1}{2}, 0)$, the condition (19) can be deduced as $n_1 = 2n$, which indicates that the normal divisor $\mathcal{T}^k = 2n\mathbf{a}_1 + n_2\mathbf{a}_2$ contains a half number of lattice points in $\mathcal{T}$. More specifically, only the translations $\mathcal{T}^k \{E | \mathbf{a}_1\}$ do not satisfy the condition (19). Thus, the translation group $\mathcal{T}$ can be decomposed into two subgroups, as

$$\mathcal{T} = \mathcal{T}^k + \mathcal{T}^k \{E | \mathbf{a}_1\} . \tag{20}$$

Consequently, the group of **k** is decomposed into 8 cosets of $\mathcal{T}^k$ after substituting (20) into (6), i.e.

$$\begin{aligned}\mathcal{G}^k &= \mathcal{T}^k \{E | \mathbf{0}\} + \mathcal{T}^k \{E | \mathbf{a}_1\} + \mathcal{T}^k \{C_2 | \mathbf{0}\} + \mathcal{T}^k \{C_2 | \mathbf{a}_1\} \\ &+ \mathcal{T}^k \{\sigma_x | \boldsymbol{\tau}\} + \mathcal{T}^k \{\sigma_x | \boldsymbol{\tau} + \mathbf{a}_1\} + \mathcal{T}^k \{\sigma_y | \boldsymbol{\tau}\} + \mathcal{T}^k \{\sigma_y | \boldsymbol{\tau} + \mathbf{a}_1\}.\end{aligned} \tag{21}$$

It is obvious that the 8 cosets in Eq. (21) are obtained by multiplying the plane group operations ($\{E|\mathbf{0}\}, \{C_2|\mathbf{0}\}, \{\sigma_x|\boldsymbol{\tau}\}, \{\sigma_y|\boldsymbol{\tau}\}$) with the translation groups $\mathcal{T}^k$ and $\mathcal{T}^k\{E|\mathbf{a}_1\}$, respectively. After tedious calculation, it can be found that the factor group $\mathcal{G}^k/\mathcal{T}^k$ is isomorphic to the point group 4$mm$, i.e. the point group of Γ point. Therefore, the small representations $D_{ji}^k$ for the $X$ point is the same as Table 2 by replacing the corresponding symmetry operations. Hence, all of the five possible small representations $X_i$ ($i=1,2,\cdots 5$) for the $X$ point are listed in Table 5. In fact, not all of the five representations $X_i$ ($i=1,2,\cdots 5$) are permissible due to the constraint condition (14). For the $X$ point, Eq. (14) can be derived as

$$D_{ji}^k(\{E|\mathbf{a}_1\}) = \exp(-i\mathbf{k}\mathbf{a}_1)\mathbf{I}_d = \exp(-i\pi)\mathbf{I}_d = -\mathbf{I}_d, \tag{22}$$

which indicates that the small representation matrix $D_{ji}^k$ corresponding to the operation $\{E|\mathbf{a}_1\}$ should be $D^k = -1$ for 1D representation or $D_{ji}^k = -\mathbf{I}$ for 2D representations. Hence from Table 5 we can conclude that only the small representation $X_5$ is permissible, whereas all the four 1D representations are spurious solutions. The analysis here also indicates that the phonon modes always have double degeneracy at $X$ point that is guaranteed by the glide symmetry.

The $Y$ point is along the line $XM$ with $\mathbf{k} = (\frac{1}{2}, \alpha)$. The analysis of $Y$ point is usually complicated so we consider a special point $\mathbf{k} = (\frac{1}{2}, \frac{1}{4})$ in this work. For $\mathbf{k} = (\frac{1}{2}, \frac{1}{4})$, the lattice translations $\mathcal{T}^k$ satisfying the condition (19) is $2n_1 + n_2 = 4n$ so we can derive the normal divisor as

$$\mathcal{T}^{\mathbf{k}} = n_1 \mathbf{a}_1 + 2(2n - n_1)\mathbf{a}_2. \tag{23}$$

It can be found from (23) that the translations in $\mathcal{T}^{\mathbf{k}}$ contain 1/4 of the lattice points in $\mathcal{T}$. These lattice points in $\mathcal{T}$ that are excluded from $\mathcal{T}^{\mathbf{k}}$ can be obtained as $\mathcal{T}^{\mathbf{k}}\{E|\mathbf{a}_1\}$, $\mathcal{T}^{\mathbf{k}}\{E|\mathbf{a}_2\}$, and $\mathcal{T}^{\mathbf{k}}\{E|-\mathbf{a}_2\}$, i.e.

$$\mathcal{T} = \mathcal{T}^{\mathbf{k}} + \mathcal{T}^{\mathbf{k}}\{E|\mathbf{a}_1\} + \mathcal{T}^{\mathbf{k}}\{E|\mathbf{a}_2\} + \mathcal{T}^{\mathbf{k}}\{E|-\mathbf{a}_2\}. \tag{24}$$

Since the point group of **k** at $Y$ point is of order 2 and there exist 4 subgroups in (24), the group $\mathcal{G}^{\mathbf{k}}$ of the point $\mathbf{k} = (\tfrac{1}{2}, \tfrac{1}{4})$ can be written in terms of 8 cosets, as

$$\begin{aligned}\mathcal{G}^{\mathbf{k}} &= \mathcal{T}^{\mathbf{k}}\{E|\mathbf{0}\} + \mathcal{T}^{\mathbf{k}}\{E|\mathbf{a}_1\} + \mathcal{T}^{\mathbf{k}}\{E|\mathbf{a}_2\} + \mathcal{T}^{\mathbf{k}}\{E|-\mathbf{a}_2\} \\ &+ \mathcal{T}^{\mathbf{k}}\{\sigma_x|\boldsymbol{\tau}\} + \mathcal{T}^{\mathbf{k}}\{\sigma_x|\boldsymbol{\tau}+\mathbf{a}_1\} + \mathcal{T}^{\mathbf{k}}\{\sigma_x|\boldsymbol{\tau}+\mathbf{a}_2\} + \mathcal{T}^{\mathbf{k}}\{\sigma_x|\boldsymbol{\tau}-\mathbf{a}_2\}.\end{aligned} \tag{25}$$

It can be verified that the factor group $\mathcal{G}^{\mathbf{k}}/\mathcal{T}^{\mathbf{k}}$ for (25) is isomorphic to the group $G_8^1$ (Bradley and Cracknell, 2010) by assuming $P = \mathcal{T}^{\mathbf{k}}\{\sigma_x|\boldsymbol{\tau}\}$ therein. Actually $G_8^1$ is the cyclic group of order 8. The irreducible representations of the factor group $\mathcal{G}^{\mathbf{k}}/\mathcal{T}^{\mathbf{k}}$ are listed in Table 6, where only 1D representations can be found. Further, the condition in Eq. (14) requires that the small representation matrices corresponding to $\{E|\mathbf{a}_1\}$, $\{E|\mathbf{a}_2\}$, and $\{E|-\mathbf{a}_2\}$ should satisfy the following three relations, as

$$\begin{aligned}D^{\mathbf{k}}(\{E|\mathbf{a}_1\}) &= \exp(-i\mathbf{k}\mathbf{a}_1) = -1, \\ D^{\mathbf{k}}(\{E|\mathbf{a}_2\}) &= \exp(-i\mathbf{k}\mathbf{a}_2) = -i, \\ D^{\mathbf{k}}(\{E|-\mathbf{a}_2\}) &= \exp(i\mathbf{k}\mathbf{a}_2) = i.\end{aligned} \tag{26}$$

After comparing Eq. (26) with Table 6, we can find that only $Y_4$ and $Y_8$ are permissible representations. Therefore, the $Y$ point at the boundary of the FBZ does not have any symmetry-guaranteed degeneracies since it only has two 1D small representations.

The $M$ point is a vertex of the FBZ with $\mathbf{k} = (\tfrac{1}{2}, \tfrac{1}{2})$. This point has the same point group symmetry as the $\Gamma$ point with 8 symmetry operations, as shown in Table 1. The analysis of the small representations to the $M$ point is similar to that for the $X$ point, although there are more symmetry operations for $M$. For $\mathbf{k} = (\tfrac{1}{2}, \tfrac{1}{2})$, the lattice translations $\mathcal{T}^{\mathbf{k}}$ satisfying the condition (19) is $n_1 + n_2 = 2n$ so that we can derive the normal divisor as

$$\mathcal{T}^{\mathbf{k}} = n_1 \mathbf{a}_1 + (2n - n_1)\mathbf{a}_2. \tag{27}$$

Further analysis indicates that $\mathcal{T}^{\mathbf{k}}$ in (27) contains a half number of the lattice points in $\mathcal{T}$. The other half of lattice translations can be recovered by multiplying $\mathcal{T}^{\mathbf{k}}$ with the translation $\{E|\mathbf{a}_1\}$, which yields the set $\mathcal{T}^{\mathbf{k}}\{E|\mathbf{a}_1\}$. Thus, the plane group $\mathcal{G}^{\mathbf{k}}$ can be decomposed into 16 cosets at the $M$ point, as

$$\begin{aligned}
\mathcal{G}^{\mathbf{k}} = &\, \mathcal{T}^{\mathbf{k}}\{E\,|\,\mathbf{0}\} + \mathcal{T}^{\mathbf{k}}\{E\,|\,\mathbf{a}_1\} + \mathcal{T}^{\mathbf{k}}\{C_2\,|\,\mathbf{0}\} + \mathcal{T}^{\mathbf{k}}\{C_2\,|\,\mathbf{a}_1\} \\
&+ \mathcal{T}^{\mathbf{k}}\{C_4^+\,|\,\mathbf{0}\} + \mathcal{T}^{\mathbf{k}}\{C_4^+\,|\,\mathbf{a}_1\} + \mathcal{T}^{\mathbf{k}}\{C_4^-\,|\,\mathbf{0}\} + \mathcal{T}^{\mathbf{k}}\{C_4^-\,|\,\mathbf{a}_1\} \\
&+ \mathcal{T}^{\mathbf{k}}\{\sigma_x\,|\,\boldsymbol{\tau}\} + \mathcal{T}^{\mathbf{k}}\{\sigma_x\,|\,\boldsymbol{\tau}+\mathbf{a}_1\} + \mathcal{T}^{\mathbf{k}}\{\sigma_y\,|\,\boldsymbol{\tau}\} + \mathcal{T}^{\mathbf{k}}\{\sigma_y\,|\,\boldsymbol{\tau}+\mathbf{a}_1\} \\
&+ \mathcal{T}^{\mathbf{k}}\{\sigma_d\,|\,\boldsymbol{\tau}\} + \mathcal{T}^{\mathbf{k}}\{\sigma_d\,|\,\boldsymbol{\tau}+\mathbf{a}_1\} + \mathcal{T}^{\mathbf{k}}\{\sigma_d'\,|\,\boldsymbol{\tau}\} + \mathcal{T}^{\mathbf{k}}\{\sigma_d'\,|\,\boldsymbol{\tau}+\mathbf{a}_1\}.
\end{aligned} \quad (28)$$

The factor group $\mathcal{G}^{\mathbf{k}}/\mathcal{T}^{\mathbf{k}}$ obtained from (28) is isomorphic to one of the point group of order 16. After classifying the cosets into different classes and test different group generators, we can eventually find that $\mathcal{G}^{\mathbf{k}}/\mathcal{T}^{\mathbf{k}}$ is isomorphic to the group $G_{16}^{10}$ (Bradley and Cracknell, 2010), which can be verified easily by taking $P = \mathcal{T}^{\mathbf{k}}\{C_4^+\,|\,\mathbf{0}\}$, $Q = \mathcal{T}^{\mathbf{k}}\{C_2\,|\,\mathbf{a}_1\}$, and $R = \mathcal{T}^{\mathbf{k}}\{\sigma_d\,|\,\boldsymbol{\tau}\}$ therein. After obtaining the group generators, other elements can be assigned readily. The irreducible representations for the factor group $\mathcal{G}^{\mathbf{k}}/\mathcal{T}^{\mathbf{k}}$ at point $M$ are listed in Table 7, which includes eight 1D representations $M_i$ $(i=1,2,\cdots 8)$ and two 2D representations $M_9$ and $M_{10}$. The representation matrices are also provided explicitly for $M_9$ and $M_{10}$ in Table 7. Furthermore, the condition in Eq. (14) requires that the small representation matrices corresponding to $\{E\,|\,\mathbf{a}_1\}$ must satisfy the following relation, as

$$\begin{aligned}
D^{\mathbf{k}}(\{E\,|\,\mathbf{a}_1\}) &= \exp(-i\mathbf{k}\mathbf{a}_1) = -1, \\
D_{ji}^{\mathbf{k}}(\{E\,|\,\mathbf{a}_1\}) &= \exp(-i\mathbf{k}\mathbf{a}_1)\mathbf{I} = -\mathbf{I}.
\end{aligned} \quad (29)$$

for 1D representation $D^{\mathbf{k}}$ and 2D representation $D_{ji}^{\mathbf{k}}$, respectively. Hence, we can infer from Table 7 that only the five representations $M_i$ $(i=5,6,7,8,9)$ satisfy the conditions in (29), among which four are 1D and one is 2D. The other five representations $M_i$ $(i=1,2,3,4,10)$ are not permissible for $M$ point. Finally, we can conclude that the phonon modes at $M$ point may have double degeneracies represented by $M_9$ that is guaranteed by the glide symmetry.

## 4 Phonon Modes at Different k-Points

### 4.1 Symmetry-Adapted Phonon Modes

The calculation procedure for phononic band structures and eigenmodes has been introduced in Section 2.3. It has been a common belief that the obtained eigenmodes are phonon modes. In fact, this is not true in special cases when the frequency degeneracy occurs. Strictly speaking, the phonon/Bloch modes should satisfy all symmetry constraints imposed by $\mathcal{G}^{\mathbf{k}}$, the group of $\mathbf{k}$, including both translation symmetry and symmetry operations related to the point group of $\mathbf{k}$. However, the Bloch condition (11) stems from the translation symmetry and won't impose any constraints to the eigenmode solutions for the latter type of symmetry. In another word, the obtained eigenmodes may not satisfy the small representations $D_{ji}^{\mathbf{k}}$.

Therefore, we present a method to derive the symmetry-adapted phonon modes from the eigenmode solutions.

The eigenmodes associated to single frequencies are discussed first. Consider an eigenmode $\phi_\mathbf{k}$ obtained from the finite element analysis in Section 2.3, which corresponds to the eigenfrequency $\omega_\mathbf{k}$. According to linear algebra theory, $\phi'_\mathbf{k} = c\phi_\mathbf{k}$ is also an eigenmode given that $c$ is a non-zero constant. The symmetry constraint imposed by Eq. (13) is

$$\{\mathbf{R}\mid\mathbf{t}\}\phi'_\mathbf{k} = \phi'_\mathbf{k} D^\mathbf{k}, \tag{30}$$

or equivalently,

$$\{\mathbf{R}\mid\mathbf{t}\}\phi_\mathbf{k} = \phi_\mathbf{k} D^\mathbf{k}. \tag{31}$$

Hence, for eigenmodes without frequency degeneracies, they satisfy the symmetry constraints automatically so there is no need to derive symmetry-adapted modes in this case.

However, the eigenmode may not be the phonon mode if degeneracies occur, which is demonstrated here for the double degeneracy case. Note that there may be two situations for the double degeneracies, either accidental degeneracy (two 1D representations) or symmetry-guaranteed degeneracy (one 2D representation). Assume that $\phi_{\mathbf{k}1}$ and $\phi_{\mathbf{k}2}$ are two eigenmodes corresponding to the same frequency $\omega_\mathbf{k}$ in the finite element analysis. Any linear combination of these two eigenmodes are also eigenmodes of the system, i.e.

$$(\phi'_{\mathbf{k}1}\ \phi'_{\mathbf{k}2}) = (\phi_{\mathbf{k}1}\ \phi_{\mathbf{k}2})\begin{bmatrix} c_{11} & c_{12} \\ c_{21} & c_{22} \end{bmatrix}, \tag{32}$$

where $c_{ij}$ ($i,j = 1,2$) are four unknown constants. Thus, there are infinite pairs of $(\phi'_{\mathbf{k}1}\ \phi'_{\mathbf{k}2})$ being candidates of the symmetry-adapted phonon modes. In general, the symmetry condition (13) requires the phonon modes $(\phi'_{\mathbf{k}1}\ \phi'_{\mathbf{k}2})$ to satisfy

$$\{\mathbf{R}\mid\mathbf{t}\}(\phi'_{\mathbf{k}1}\ \phi'_{\mathbf{k}2}) = (\phi'_{\mathbf{k}1}\ \phi'_{\mathbf{k}2})\begin{bmatrix} D^\mathbf{k}_{11} & D^\mathbf{k}_{12} \\ D^\mathbf{k}_{21} & D^\mathbf{k}_{22} \end{bmatrix}. \tag{33}$$

The $D^\mathbf{k}_{ij}$ in (33) is a 2D representation matrix for cases with symmetry-guaranteed degeneracies. However if accidental degeneracy occurs, $D^\mathbf{k}_{ij}$ consists of two 1D representations $D^\mathbf{k}_{11}$ and $D^\mathbf{k}_{22}$ while the other two terms vanish as $D^\mathbf{k}_{12} = D^\mathbf{k}_{21} = 0$. In general cases, after substituting (32) into (33), it yields

$$\{\mathbf{R}|\mathbf{t}\}(\phi_{\mathbf{k}1}\ \phi_{\mathbf{k}2}) = (\phi_{\mathbf{k}1}\ \phi_{\mathbf{k}2})\begin{bmatrix} c_{11} & c_{12} \\ c_{21} & c_{22} \end{bmatrix}\begin{bmatrix} D^{\mathbf{k}}_{11} & D^{\mathbf{k}}_{12} \\ D^{\mathbf{k}}_{21} & D^{\mathbf{k}}_{22} \end{bmatrix}\begin{bmatrix} c_{11} & c_{12} \\ c_{21} & c_{22} \end{bmatrix}^{-1}. \qquad (34)$$

Thus, the four coefficients $c_{ij}$ ($i, j = 1, 2$) in (32) can be obtained by applying the symmetry condition (34) for the eigenmode solution $(\phi_{\mathbf{k}1}\ \phi_{\mathbf{k}2})$ from finite element analysis. Since the amplitude of the phonon modes $(\phi'_{\mathbf{k}1}\ \phi'_{\mathbf{k}2})$ does not matter, only three constants in $c_{ij}$ ($i, j = 1, 2$) need to be solved if $D^{\mathbf{k}}_{ij}$ is a 2D representation (symmetry-guaranteed degeneracy), while only two constants of $c_{ij}$ ($i, j = 1, 2$) are needed if $D^{\mathbf{k}}_{ij}$ indicates two 1D representations (accidental degeneracy).

For the frequency degeneracies on order 3 or higher, similar technique can be used to determine the symmetry-adapted phonon modes, which will not be introduced here since they have not been found for the current *p*4*g* structure. Note that all the phonon modes presented in this work are symmetry-adapted phonon modes.

Actually, researchers working on high-frequency homogenization of periodic materials have realized the degeneracies at some **k**-points (Harutyunyan et al., 2016) and developed an alternative method (Guzina et al., 2018) to decouple the phonon modes when multiple degeneracy exists. However, the phonon modes obtained by using their method may not satisfy the symmetry relations outlined in this work since the bases of the small representations can be transformed.

### 4.2  Phononic Band Structure

The phononic band structure of the considered *p*4*g* lattice structure is obtained for the contour $\Gamma$-$\Delta$-$X$-$Y$-$M$-$\Sigma$-$\Gamma$ in the irreducible FBZ. The geometry and material properties of the lattice structure have been introduced in Section 2.3. Figure 4 illustrates the first 8 bands for each high symmetry point. The group theory requires that each band is associated to a small representation $D^{\mathbf{k}}_{ji}$ of the corresponding **k**-point. Hence besides the dispersion curves, we also label the small representation for each band. There are at least five reasons to do it.

(1) *Phonon mode symmetry*. Symmetry is a fundamental property of phonon modes. Uncovering the symmetry of phonon modes will not only help understand the dynamic characteristics of periodic structures but also facilitate the design and application of phononic device for wave manipulating usage at certain frequencies or along specific directions (Celli and Gonella, 2015). In addition, the symmetry of phonon modes is also related to the dynamic response spectra of periodic structures.

(2) *Frequency degeneracy*. Labeling the representation of bands can clarify the frequency degeneracies at some **k**-points. Even though only single frequencies are found at most **k**-points, double degeneracies also occur in the band structure shown in Figure 4. For example, the bands appear as coalescing pairs along the FBZ boundary *X*-*Y*-*M*, which is called the sticking-bands effect as a typical feature of nonsymmorphic groups. In addition, double degeneracies also occur for the $\Gamma$ point and some of the $\Sigma$ points. These double degeneracies may be caused by two reasons, either due to 2D representations or accidental degeneracy. Identifying the representation of each band will distinguish these degeneracies.

(3) *Band crossing*. Labeling the bands also clarify whether band crossing or anticrossing occurs at some of the **k**-points since the group theory demands that bands of the same representation should not cross each other. This is useful for band structure calculation since one does not need to compute a large amount of **k**-points near the crossing points to identify the trajectory of bands. This method is also useful when the distance between two repulsed bands is so small that even refining the **k**-points may not yield correct answer (Lu and Srivastava, 2018).

(4) *Band sorting*. A band structure usually consists of multiple bands crossing each other (Lu and Srivastava, 2018). It is rather impossible to sort the sequence of these bands properly without identifying the representation of each band. In theory, the representation is continuous along a band regardless of the crossing point. Hence, this is a powerful tool to sort the bands when multiple crossing points exist.

(5) *Band gaps*. First and for most, the representations of bands will help identify narrow band gaps since they are easy to be ignored. Moreover, researchers also found that the glide symmetry may have even more significant implications to the design of phononic crystals. For example, it is believed that introducing glide symmetry may broaden the band gaps since some of the bands stick together and the curvature of bands decrease (Gan, 2017; Koh, 2011).

**4.3   Symmetry of Phonon Modes**

The symmetry of phonon modes for each band is discussed in this section. For the cases with frequency degeneracies we will use the symmetry-adapted phonon modes derived in Section 4.1. Four primitive unit cells are depicted to show both the real and imaginary parts of the phonon modes. The origin of the coordinate system is set as the center of the lower left unit cell.

At the $\Gamma$ point, both 1D and 2D representations are found. Since the $\Delta$ and $\Sigma$ points only exhibit 1D representations, each of the 1D representations $\Gamma_i$ ($i=1,2,3,4$) should connect to a single band at the $\Delta$ or $\Sigma$ point, while the only 2D representation $\Gamma_5$ is connected by two bands. Thus, one can readily identify the frequencies with $\Gamma_5$ representation. Of course, the phonon modes for $\Gamma_5$ must be further verified by the symmetry operations in Table 2 since the accidental degeneracy may occur sometimes. The phonon modes of 1D representation can be determined by examining each phonon mode with the symmetry conditions in Table 2, e.g. $\{C_4^+|\mathbf{0}\}$ and $\{\sigma_x|\boldsymbol{\tau}\}$ are adequate to identify the corresponding 1D representation. For the first 8 bands, only the 1D representations $\Gamma_2$ and $\Gamma_3$ are found, while the other two 1D representations may appear for higher order bands. Figure 5 illustrates the phonon modes represented by the two 1D representations ($\Gamma_2$, $\Gamma_3$) and the 2D representation $\Gamma_5$. The angular frequencies for these phonon modes are chosen as $\bar{\omega} =$ 94.96, 150.10, and 63.64 for $\Gamma_2$, $\Gamma_3$, and $\Gamma_5$, respectively. The 2D representation $\Gamma_5$ corresponds to two phonon modes $\phi_{\mathbf{k}1}$ and $\phi_{\mathbf{k}2}$ as shown in Figure 5. Note that we have used the symmetry-adapted phonon modes here and the four constants $c_{ij}$ can be determined from the symmetry conditions (34) for operations $\{C_4^+ \mid \mathbf{0}\}$ and $\{\sigma_x \mid \boldsymbol{\tau}\}$.

The $\Delta$ point is relatively easy to analyze since only two 1D representations exist, i.e. $\Delta_1$ and $\Delta_2$. The only symmetry operation needs to be examined is $\{\sigma_y \mid \boldsymbol{\tau}\}$ in Table 3. The symmetry condition for each of the representations is $\{\sigma_y \mid \boldsymbol{\tau}\}\phi_{\mathbf{k}} = \pm\exp(-\alpha\pi i)\phi_{\mathbf{k}}$, with the positive sign for $\Delta_1$ and negative sign for the other. After examining the symmetry condition for the **k**-points of interest, we can determine the representation of each band as shown in Figure 4. There are two compatibility rules helping determine the representations (Sakoda, 2005). At first, the representation $\Gamma_2$ is compatible with $\Delta_2$ while $\Gamma_3$ is compatible with $\Delta_1$. So that the bands connecting $\Gamma_2$ and $\Gamma_3$ can be determined easily. Secondly, the representation $\Gamma_5$ must be connected by two distinct bands $\Delta_1$ and $\Delta_2$. Hence the representations of the two bands are clear once one of them is determined. Another interesting phenomenon observed at the $\Delta$ point is the anticrossing between the 6$^{th}$ and 7$^{th}$ bands. It can be found that both of these two bands are represented by $\Delta_1$, indicating that these two bands must be anticrossing. This feature is very convenient when one generates the dispersion curves from discrete data points since there is no need to refine the **k**-points to determine whether crossing or anticrossing occurs. Typical phonon modes for $\Delta_1$ and $\Delta_2$ are illustrated in Figure 6, where the $\Delta_1$ mode is taken from the 1$^{st}$ band and the $\Delta_2$ mode is from the 2$^{nd}$ band with $\mathbf{k} = (\frac{1}{4}, 0)$ for both modes.

The analysis of the $\Sigma$ point is similar to that for $\Delta$. There are only two 1D representations $\Sigma_1$ and $\Sigma_2$ so the symmetry of each phonon mode can be determined by examining the symmetry condition $\{\sigma_d' \mid \boldsymbol{\tau}\}\phi_{\mathbf{k}} = \pm\exp(-2\alpha\pi i)\phi_{\mathbf{k}}$ in Table 4. Among the first 8 bands, the 1$^{st}$, 4$^{th}$, 5$^{th}$, and 8$^{th}$ bands do not cross with any other bands so their representations can be determined first. However, the remaining four bands are more complicated to analyze since they have crossing points. There are four branches near each of the two crossing points so we need to determine the representation for all branches. Again, the compatibility rules can be utilized. It is found from Table 2 and Table 4 that $\Gamma_2$ and $\Gamma_3$ are compatible to $\Sigma_2$ only. In addition, the 2D representation $\Gamma_5$ should be connected by two distinct bands $\Sigma_1$ and $\Sigma_2$. These rules help determine the representations of some of the bands. Other bands need to be examined by the symmetry condition. For all the bands at the $\Sigma$ point, band crossing between the $\Sigma_1$ and $\Sigma_2$ bands is observed. Identifying the crossing points is significant for the band sorting purpose. Figure 6 illustrates two typical phonon modes for the $\Sigma_1$ and $\Sigma_2$ bands, where the $\Sigma_1$ and $\Sigma_2$ modes are taken for $\bar{\omega} = 37.87$ and $\bar{\omega} = 8.03$, respectively, both at the point $\mathbf{k} = (\frac{1}{4}, \frac{1}{4})$.

For the $X$ point at the boundary of the FBZ, only one 2D representation $X_5$ is permissible so it is very easy to label the modes. This also implies that the frequency degeneracy is induced by the 2D representation at $X$ rather than accidental degeneracy. The 2D representation $X_5$ is connected to two distinct 1D representations $\Delta_1$ and $\Delta_2$ at the $\Delta$ point according to the compatibility rule. Two phonon modes $\phi_{\mathbf{k}1}$ and $\phi_{\mathbf{k}2}$ are illustrated In Figure 7 for the $X$ point, which are drawn from the frequency $\bar{\omega} = 31.42$. The four constants $c_{ij}$ required by the symmetry-adapted modes are obtained from the symmetry conditions (34) for $\{\sigma_y | \boldsymbol{\tau}\}$ and $\{C_2 | \mathbf{0}\}$. For all the phonon modes at the $X$ point, $\phi_{\mathbf{k}}$ is always periodic along the $y$ direction since $\{E | \mathbf{a}_2\}\phi_{\mathbf{k}} = \phi_{\mathbf{k}}$. However, along the $x$ direction, a lattice translation yields the negative phonon mode $\{E | \mathbf{a}_1\}\phi_{\mathbf{k}} = -\phi_{\mathbf{k}}$ while another lattice translation brings back the original phonon mode $\{E | 2\mathbf{a}_1\}\phi_{\mathbf{k}} = \phi_{\mathbf{k}}$.

Similar to the $X$ point, all the $Y$ points also have double degeneracies for the frequencies, which are shown in Figure 4. However, unlike the $X$ point with a 2D representation, the $Y$ points only have two permissible 1D representations, indicating that the frequency degeneracy is accidental instead of symmetry-guaranteed. Each of the $X_5$ representation is connected to two distinct 1D representations according to the compatibility rule. Thus for the $Y$ point, the $Y_4$ and $Y_8$ bands always overlap with each other and appear as pairs. In Figure 7 we illustrate two phonon modes with frequency $\bar{\omega} = 24.81$ at $\mathbf{k} = (\frac{1}{2}, \frac{1}{4})$. The four constants $c_{ij}$ generating the symmetry-adapted modes are obtained by applying the symmetry condition (34) for the operation $\{\sigma_x | \boldsymbol{\tau}\}$.

The symmetry of phonon modes at the $M$ point is relatively more complicated than other $\mathbf{k}$-points. It is found from Figure 4 that double degeneracy occurs for all frequencies. However, it is not clear whether each degeneracy is induced by accidence or the symmetry since there exist four 1D representations and one 2D representation at the $M$ point. Further analysis of the band structure reveals that the frequencies $\bar{\omega} = 14.86$ and $\bar{\omega} = 168.57$ always connect to two $\Sigma_2$ bands, indicating that they must be the 1D representations $M_7$ and $M_8$ according to the compatibility rule. Similarly, the frequency $\bar{\omega} = 52.19$ is connected to two $\Sigma_1$ bands and hence the two phonon modes are $M_5$ and $M_6$. At last, the frequency $\bar{\omega} = 118.80$ is connected to $\Sigma_1$ and $\Sigma_2$ so it is possible the 2D representation $M_9$, which can be verified by examining the symmetry of the modes. Therefore, all the representations for the $M$ point are labeled in Figure 4. Typical phonon modes for these five representations are illustrated in Figure 8, where the modes $M_5$ and $M_6$ are taken at the frequency $\bar{\omega} = 52.19$, the modes $M_5$ and $M_6$ are taken at $\bar{\omega} = 52.19$, and the two modes for $M_9$ are taken at $\bar{\omega} = 118.80$. Since frequency degeneracies always occur, we need to determine the four constants $c_{ij}$ to generate the symmetry-adapted modes. The symmetry condition (34) is applied for the operation $\{C_4^+ | \mathbf{0}\}$ to obtain $c_{ij}$ for the four 1D representations $M_i$ $(i = 1, 2, 3, 4)$, while two operations $\{\sigma_x | \boldsymbol{\tau}\}$ and $\{C_4^+ | \mathbf{0}\}$ are needed to determine $c_{ij}$ for the 2D representation $M_9$.

## 5 Symmetry of Periodic Bloch Functions for Interior k-Points

The symmetry of phonon modes is usually considered for the Bloch waves $\phi_\mathbf{k} = (u_x \ u_y)_\mathbf{k}$, while the symmetry of their periodic term $\tilde{\phi}_\mathbf{k} = (\tilde{u}_x \ \tilde{u}_y)_\mathbf{k}$ is seldom explored. In fact, the small representations based on the periodic Bloch functions $\tilde{\phi}_\mathbf{k} = (\tilde{u}_x \ \tilde{u}_y)_\mathbf{k}$ may exhibit simpler mathematical forms than that for the Bloch waves $\phi_\mathbf{k} = (u_x \ u_y)_\mathbf{k}$, at least for all the interior **k**-points within the FBZ. This simplification will actually alleviate the complexity in determining the representation for each mode.

The symmetry of periodic Bloch functions $\tilde{\phi}_\mathbf{k} = (\tilde{u}_x \ \tilde{u}_y)_\mathbf{k}$ for interior **k**-points is derived in what follows. Similar to Eq. (9), we can decouple the Bloch wave $\phi_\mathbf{k} = (u_x \ u_y)_\mathbf{k}$ into two terms, as

$$\phi_\mathbf{k}(\mathbf{x}) = \tilde{\phi}_\mathbf{k}(\mathbf{x}) e^{i\mathbf{k}\mathbf{x}} \ , \tag{35}$$

where $\tilde{\phi}_\mathbf{k} = (u_x \ u_y)_\mathbf{k}$. Substituting Eq. (35) into Eq. (13) directly will complicate the latter even further due to the existence of the term $e^{i\mathbf{k}\mathbf{x}}$. Nonetheless, this is not the case for Eq. (16). After the substitution of Eq. (35) into Eq. (16), it yields

$$\{\mathbf{R} | \mathbf{t}\}(\tilde{\phi}_\mathbf{k} e^{i\mathbf{k}\mathbf{x}}) = e^{i\mathbf{k}(\mathbf{x}-\mathbf{t})} \sum_{j=1}^{d} \tilde{\phi}_\mathbf{k} D_{ji}(\mathbf{R}) \ . \tag{36}$$

The term on the left hand side of (36) is further derived as

$$\begin{aligned} \{\mathbf{R} | \mathbf{t}\}(\tilde{\phi}_\mathbf{k} e^{i\mathbf{k}\mathbf{x}}) &= \{\mathbf{R} | \mathbf{t}\}\tilde{\phi}_\mathbf{k} \cdot \{\mathbf{R} | \mathbf{t}\} e^{i\mathbf{k}\mathbf{x}} \\ &= \{\mathbf{R} | \mathbf{t}\}\tilde{\phi}_\mathbf{k} \cdot \exp(i\mathbf{k}\{\mathbf{R} | \mathbf{t}\}^{-1}\mathbf{x}) \\ &= \{\mathbf{R} | \mathbf{t}\}\tilde{\phi}_\mathbf{k} \cdot \exp(i\mathbf{k}\mathbf{R}^{-1}(\mathbf{x}-\mathbf{t})) \ , \\ &= \{\mathbf{R} | \mathbf{t}\}\tilde{\phi}_\mathbf{k} \cdot e^{i\mathbf{R}\mathbf{k}(\mathbf{x}-\mathbf{t})} \\ &= \{\mathbf{R} | \mathbf{t}\}\tilde{\phi}_\mathbf{k} \cdot e^{i\mathbf{k}(\mathbf{x}-\mathbf{t})} \end{aligned} \tag{37}$$

Note that we have used the fact that $\mathbf{Rk} = \mathbf{k}$ for all the interior **k**-points when $\{\mathbf{R}|\mathbf{t}\}$ is in $\mathcal{G}^{\mathbf{k}}$. By comparing Eq. (36) and Eq. (37), we can readily conclude that for all the interior **k**-points,

$$\{\mathbf{R}|\mathbf{t}\}\tilde{\phi}_{\mathbf{k}} = \sum_{j}\tilde{\phi}_{\mathbf{k}} D_{ji}(\mathbf{R}) . \tag{38}$$

Therefore, one can eliminate the phase factor $\exp(-i\mathbf{kt})$ in Eq. (16) if the periodic Bloch functions $\tilde{\phi}_{\mathbf{k}} = (\tilde{u}_x \ \tilde{u}_y)_{\mathbf{k}}$ are introduced. The small representation in Eq. (38) is very easy to use since the table of $D_{ji}(\mathbf{R})$ can be found easily. It is emphasized here that the simplification is merely for all the interior **k**-points. For any boundary **k**-points, Eq. (38) may not be valid anymore since the relation $\mathbf{Rk} = \mathbf{k} + \mathbf{b}_n$ holds, where $\mathbf{b}_n$ is an arbitrary reciprocal lattice translation vector. In this case, the phase factor won't be eliminated in Eq. (37) and there is no gain from introducing the periodic Bloch functions.

The $\Gamma$ point is not introduced here anymore since the phonon modes are identical to the periodic Bloch functions in this case, i.e. $\tilde{\phi}_{\mathbf{k}} = \phi_{\mathbf{k}}$ for $\mathbf{k} = \mathbf{0}$. The small representations $D_{ji}$ for the $\Delta$ and $\Sigma$ points are given in Table 8 and Table 9, respectively, if the periodic Bloch functions $\tilde{\phi}_{\mathbf{k}}$ are chosen as the bases. In fact, Table 8 and Table 9 can be derived from Table 3 and Table 4 by omitting the phase factor $\exp(-i\mathbf{kt})$. Due to the simplification, it becomes easier to determine the symmetry of each phonon mode by examining the symmetry of its periodic Bloch function. The periodic Bloch functions $\tilde{\phi}_{\mathbf{k}}$ corresponding to the phonon modes in Figure 6 are presented in Figure 9. The symmetry condition (38) can be verified for the modes in Figure 9 readily by using the $D_{ji}$ shown in Table 8 and Table 9.

## 6 Conclusions

The phononic band structure of periodic structures with glide symmetry usually exhibits frequency degeneracies at the boundary of the FBZ. This phenomenon is called the sticking-bands effect as a typical feature of nonsymmorphic structures. Elucidating the symmetry of phonon modes and the frequency degeneracies will not only help the band sorting by identifying the band crossing and anticrossing correctly but also facilitate the design and application of these periodic structures. However, the symmetry analysis is usually very challenging since it involves the small representation of nonsymmorphic space/plane groups due to the existence of glide symmetry. To address this issue, this work considers a *p*4*g* lattice structure as an example to show the general procedure determining the symmetry and degeneracies of phonon modes for periodic structures with glide symmetry. Different techniques are needed for the **k**-points inside the FBZ and those at the boundary, where the Herring's method is used.

Periodic structures in the *p*4*g* group have the highest symmetry among the four nonsymmorphic plane groups. A large family of phononic crystals and mechanical metamaterials fall into the *p*4*g* group. In this work, the phononic band structure of the *p*4*g* lattice structure is obtained for the high symmetry **k**-points in the irreducible FBZ. More importantly, the small representations are tabulated for all the high symmetry **k**-points to analyze the symmetry of the phonon modes. Based on the theory and analysis, the small representation of each phononic band is labeled and typical symmetry-adapted phonon modes are illustrated. It has been found that the frequency degeneracies at the boundary of FBZ can be classified into two types: symmetry-guaranteed degeneracy and accidental degeneracy. The double degeneracies at the *X* point belong to the former type and only one 2D representation $X_5$ exists. The frequency degeneracies at the *Y* point belong to the latter type and two 1D representations $Y_4$ and $Y_8$ appear as pairs. As to the *M* point, both symmetry-guaranteed degeneracy and accidental degeneracy exist for different frequencies and must be analyzed with caution. In addition, the small representations based on the periodic Bloch function bases are derived for interior **k**-points, which show simplified mathematical form by eliminating the phase factors and are more convenient to determine the symmetry of phonon modes compared to the original phonon mode bases.

It is noted that the small representations presented in this work are valid for all periodic structures in the *p*4*g* group, not only the sinusoidal lattice. In addition, the method introduced in this work can also be applied to other periodic structures with/out glide symmetry. Actually there are a lot more line/plane/space groups with glide or screw symmetry to be explored in the future. Analyzing the fundamental symmetry behavior of these groups will facilitate the design of novel lattice structures, phononic crystals, wave guides, metamaterials, and among others.

It is worth to mention that the results presented in this work can be applied to the model reduction and bifurcation analysis of structures. The model reduction of structures (Maurin et al., 2017; Zingoni, 2009) usually employs the symmetry to reduce the computational cost and obtain correct results when bifurcation occurs. In addition, the bifurcation analysis of periodic structures (Combescure et al., 2016) involves identifying the buckling modes by using the representations of the corresponding space groups. Therefore, the small representations derived in this work are expected to facilitate the advanced symmetry-based analysis of periodic structures with glide symmetry.

## 7 Acknowledgements

The author acknowledges the funding support from the Watson School of Engineering and Applied Science at Binghamton University.

# References

Åberg, M., Gudmundson, P., 1997. The usage of standard finite element codes for computation of dispersion relations in materials with periodic microstructure. J. Acoust. Soc. Am. 102, 2007–2013. https://doi.org/10.1121/1.419652

Alagappan, G., Sun, X.W., Sun, H.D., 2008. Symmetries of the eigenstates in an anisotropic photonic crystal. Phys. Rev. B 77. https://doi.org/10.1103/PhysRevB.77.195117

Alderson, A., Alderson, K.L., Attard, D., Evans, K.E., Gatt, R., Grima, J.N., Miller, W., Ravirala, N., Smith, C.W., Zied, K., 2010. Elastic constants of 3-, 4- and 6-connected chiral and anti-chiral honeycombs subject to uniaxial in-plane loading. Compos. Sci. Technol. 70, 1042–1048. https://doi.org/10.1016/j.compscitech.2009.07.009

Altmann, S.L., 2002. Band theory of solids: an introduction from the point of view of symmetry, 1. publ., reprint. ed, Oxford science publications. Clarendon Press, Oxford.

Altmann, S.L., Herzig, P., 1994. Point-group theory tables. Clarendon Press ; Oxford University Press, Oxford : New York.

Aroyo, M.I. (Ed.), 2016. International tables for crystallography. Volume A: Space-group symmetry, Sixth, revised edition. ed. Wiley, Chichester, West Sussex.

Bertoldi, K., Boyce, M.C., 2008. Wave propagation and instabilities in monolithic and periodically structured elastomeric materials undergoing large deformations. Phys. Rev. B 78. https://doi.org/10.1103/PhysRevB.78.184107

Bertoldi, K., Boyce, M.C., Deschanel, S., Prange, S.M., Mullin, T., 2008. Mechanics of deformation-triggered pattern transformations and superelastic behavior in periodic elastomeric structures. J. Mech. Phys. Solids 56, 2642–2668. https://doi.org/10.1016/j.jmps.2008.03.006

Bertoldi, K., Reis, P.M., Willshaw, S., Mullin, T., 2010. Negative Poisson's Ratio Behavior Induced by an Elastic Instability. Adv. Mater. 22, 361–366. https://doi.org/10.1002/adma.200901956

Bertoldi, K., Vitelli, V., Christensen, J., van Hecke, M., 2017. Flexible mechanical metamaterials. Nat. Rev. Mater. 2, 17066. https://doi.org/10.1038/natrevmats.2017.66

Bradley, C.J., Cracknell, A.P., 2010. The mathematical theory of symmetry in solids: representation theory for point groups and space groups, Oxford classic texts in the physical sciences. Clarendon Press, Oxford.

Celli, P., Gonella, S., 2015. Tunable directivity in metamaterials with reconfigurable cell symmetry. Appl. Phys. Lett. 106, 091905. https://doi.org/10.1063/1.4914011

Chen, Y., Li, T., Scarpa, F., Wang, L., 2017a. Lattice Metamaterials with Mechanically Tunable Poisson's Ratio for Vibration Control. Phys. Rev. Appl. 7. https://doi.org/10.1103/PhysRevApplied.7.024012

Chen, Y., Qian, F., Zuo, L., Scarpa, F., Wang, L., 2017b. Broadband and multiband vibration mitigation in lattice metamaterials with sinusoidally-shaped ligaments. Extreme Mech. Lett. 17, 24–32. https://doi.org/10.1016/j.eml.2017.09.012

Cho, Y., Shin, J.-H., Costa, A., Kim, T.A., Kunin, V., Li, J., Lee, S.Y., Yang, S., Han, H.N., Choi, I.-S., Srolovitz, D.J., 2014. Engineering the shape and structure of materials by fractal cut. Proc. Natl. Acad. Sci. 111, 17390–17395. https://doi.org/10.1073/pnas.1417276111

Clausen, A., Wang, F., Jensen, J.S., Sigmund, O., Lewis, J.A., 2015. Topology Optimized Architectures with Programmable Poisson's Ratio over Large Deformations. Adv. Mater. 27, 5523–5527. https://doi.org/10.1002/adma.201502485

Combescure, C., Henry, P., Elliott, R.S., 2016. Post-bifurcation and stability of a finitely strained hexagonal honeycomb subjected to equi-biaxial in-plane loading. Int. J. Solids Struct. 88–89, 296–318. https://doi.org/10.1016/j.ijsolstr.2016.02.016

Coulais, C., 2016. Periodic cellular materials with nonlinear elastic homogenized stress-strain response at small strains. Int. J. Solids Struct. 97–98, 226–238. https://doi.org/10.1016/j.ijsolstr.2016.07.025

Dresselhaus, M., Dresselhaus, G., Jorio, A., 2008. Group Theory: Application to the Physics of Condensed Matter. Springer Berlin Heidelberg.

Dudek, K.K., Gatt, R., Mizzi, L., Dudek, M.R., Attard, D., Evans, K.E., Grima, J.N., 2017. On the dynamics and control of mechanical properties of hierarchical rotating rigid unit auxetics. Sci. Rep. 7, 46529. https://doi.org/10.1038/srep46529

Florijn, B., Coulais, C., van Hecke, M., 2014. Programmable Mechanical Metamaterials. Phys. Rev. Lett. 113. https://doi.org/10.1103/PhysRevLett.113.175503
Gan, W.S., 2017. New acoustics based on metamaterials. Springer Berlin Heidelberg, New York, NY.
Grima, J.N., Alderson, A., Evans, K.E., 2005. Auxetic behaviour from rotating rigid units. Phys. Status Solidi B 242, 561–575. https://doi.org/10.1002/pssb.200460376
Grima, J.N., Evans, K.E., 2000. Auxetic behavior from rotating squares. J. Mater. Sci. Lett. 19, 1563–1565. https://doi.org/10.1023/A:1006781224002
Grima, J.N., Gatt, R., Farrugia, P.-S., 2008. On the properties of auxetic meta-tetrachiral structures. Phys. Status Solidi B 245, 511–520. https://doi.org/10.1002/pssb.200777704
Guzina, B., Meng, S., Oudghiri-Idrissi, O., 2018. A rational framework for dynamic homogenization at finite wavelengths and frequencies. ArXiv180507496 Math.
Haghpanah, B., Papadopoulos, J., Mousanezhad, D., Nayeb-Hashemi, H., Vaziri, A., 2014. Buckling of regular, chiral and hierarchical honeycombs under a general macroscopic stress state. Proc. R. Soc. Math. Phys. Eng. Sci. 470, 20130856–20130856. https://doi.org/10.1098/rspa.2013.0856
Harutyunyan, D., Milton, G.W., Craster, R.V., 2016. High-frequency homogenization for travelling waves in periodic media. Proc. R. Soc. Math. Phys. Eng. Sci. 472, 20160066. https://doi.org/10.1098/rspa.2016.0066
Hergert, W., Däne, M., 2003. Group theoretical investigations of photonic band structures. Phys. Status Solidi A 197, 620–634. https://doi.org/10.1002/pssa.200303110
Hussein, M.I., Leamy, M.J., Ruzzene, M., 2014. Dynamics of Phononic Materials and Structures: Historical Origins, Recent Progress, and Future Outlook. Appl. Mech. Rev. 66, 040802. https://doi.org/10.1115/1.4026911
Inui, T., Tanabe, Y., Onodera, Y., 1990. Group Theory and Its Applications in Physics. Springer Berlin Heidelberg, Berlin, Heidelberg.
Jacobs, P.W.M., 2005. Group theory with applications in chemical physics. Cambridge University Press, Cambridge.
Javid, F., Wang, P., Shanian, A., Bertoldi, K., 2016. Architected Materials with Ultra-Low Porosity for Vibration Control. Adv. Mater. 28, 5943–5948. https://doi.org/10.1002/adma.201600052
Kerszberg, N., Suryanarayana, P., 2015. Ab initio strain engineering of graphene: opening bandgaps up to 1 eV. RSC Adv. 5, 43810–43814. https://doi.org/10.1039/C5RA03422A
Koh, C.Y., 2011. Generalized phononic networks : of length scales, symmetry breaking and (non) locality : "controlling complexity through simplicity" (PhD Dissertation). Massachusetts Institute of Technology.
Körner, C., Liebold-Ribeiro, Y., 2015. A systematic approach to identify cellular auxetic materials. Smart Mater. Struct. 24, 025013. https://doi.org/10.1088/0964-1726/24/2/025013
Lakes, R.S., 2017. Negative-Poisson's-Ratio Materials: Auxetic Solids. Annu. Rev. Mater. Res. 47, 63–81. https://doi.org/10.1146/annurev-matsci-070616-124118
Laude, V., 2015. Phononic crystals: artificial crystals for sonic, acoustic, and elastic waves, De Gruyter studies in mathematical physics. De Gruyter, Berlin.
Lee, J.-H., Singer, J.P., Thomas, E.L., 2012. Micro-/Nanostructured Mechanical Metamaterials. Adv. Mater. 24, 4782–4810. https://doi.org/10.1002/adma.201201644
Liu, J., Gu, T., Shan, S., Kang, S.H., Weaver, J.C., Bertoldi, K., 2016. Harnessing Buckling to Design Architected Materials that Exhibit Effective Negative Swelling. Adv. Mater. 28, 6619–6624. https://doi.org/10.1002/adma.201600812
Lu, Y., Srivastava, A., 2018. Level repulsion and band sorting in phononic crystals. J. Mech. Phys. Solids 111, 100–112. https://doi.org/10.1016/j.jmps.2017.10.021
Lustig, B., Shmuel, G., 2018. On the band gap universality of multiphase laminates and its applications. J. Mech. Phys. Solids 117, 37–53. https://doi.org/10.1016/j.jmps.2018.04.008
Maurin, F., Claeys, C., Deckers, E., Desmet, W., 2018. Probability that a band-gap extremum is located on the irreducible Brillouin-zone contour for the 17 different plane crystallographic lattices. Int. J. Solids Struct. 135, 26–36. https://doi.org/10.1016/j.ijsolstr.2017.11.006
Maurin, F., Claeys, C., Van Belle, L., Desmet, W., 2017. Bloch theorem with revised boundary conditions applied to glide, screw and rotational symmetric structures. Comput. Methods Appl. Mech. Eng. 318, 497–513. https://doi.org/10.1016/j.cma.2017.01.034

Maurin, F., Spadoni, A., 2014. Wave dispersion in periodic post-buckled structures. J. Sound Vib. 333, 4562–4578. https://doi.org/10.1016/j.jsv.2014.04.029

Mock, A., Lu, L., O'Brien, J., 2010. Space group theory and Fourier space analysis of two-dimensional photonic crystal waveguides. Phys. Rev. B 81. https://doi.org/10.1103/PhysRevB.81.155115

Mullin, T., Deschanel, S., Bertoldi, K., Boyce, M.C., 2007. Pattern Transformation Triggered by Deformation. Phys. Rev. Lett. 99. https://doi.org/10.1103/PhysRevLett.99.084301

Nanda, A., Karami, M.A., 2018. Tunable bandgaps in a deployable metamaterial. J. Sound Vib. 424, 120–136. https://doi.org/10.1016/j.jsv.2018.03.015

Ohno, N., Okumura, D., Noguchi, H., 2002. Microscopic symmetric bifurcation condition of cellular solids based on a homogenization theory of finite deformation. J. Mech. Phys. Solids 50, 1125–1153. https://doi.org/10.1016/S0022-5096(01)00106-5

Phani, A.S., Hussein, M.I., 2017. Dynamics of lattice materials.

Pratapa, P.P., Suryanarayana, P., Paulino, G.H., 2018. Bloch wave framework for structures with nonlocal interactions: Application to the design of origami acoustic metamaterials. J. Mech. Phys. Solids 118, 115–132. https://doi.org/10.1016/j.jmps.2018.05.012

Saiki, I., Ikeda, K., Murota, K., 2005. Flower patterns appearing on a honeycomb structure and their bifurcation mechanism. Int. J. Bifurc. Chaos 15, 497–515. https://doi.org/10.1142/S021812740501217X

Sakoda, K., 2005. Optical Properties of Photonic Crystals. Springer, Berlin; Heidelberg.

Saxena, K.K., Das, R., Calius, E.P., 2016. Three Decades of Auxetics Research − Materials with Negative Poisson's Ratio: A Review: Three Decades of Auxetics Research…. Adv. Eng. Mater. 18, 1847–1870. https://doi.org/10.1002/adem.201600053

Shim, J., Shan, S., Košmrlj, A., Kang, S.H., Chen, E.R., Weaver, J.C., Bertoldi, K., 2013. Harnessing instabilities for design of soft reconfigurable auxetic/chiral materials. Soft Matter 9, 8198. https://doi.org/10.1039/c3sm51148k

Sigmund, O., Torquato, S., Aksay, I.A., 1998. On the design of 1–3 piezocomposites using topology optimization. J. Mater. Res. 13, 1038–1048. https://doi.org/10.1557/JMR.1998.0145

Simulia, 2017. Abaqus Analysis User's Guide.

Tang, Y., Lin, G., Han, L., Qiu, S., Yang, S., Yin, J., 2015. Design of Hierarchically Cut Hinges for Highly Stretchable and Reconfigurable Metamaterials with Enhanced Strength. Adv. Mater. 27, 7181–7190. https://doi.org/10.1002/adma.201502559

Trainiti, G., Rimoli, J.J., Ruzzene, M., 2016. Wave propagation in undulated structural lattices. Int. J. Solids Struct. 97–98, 431–444. https://doi.org/10.1016/j.ijsolstr.2016.07.006

Trainiti, G., Rimoli, J.J., Ruzzene, M., 2015. Wave propagation in periodically undulated beams and plates. Int. J. Solids Struct. 75–76, 260–276. https://doi.org/10.1016/j.ijsolstr.2015.08.019

Yu, X., Zhou, J., Liang, H., Jiang, Z., Wu, L., 2018. Mechanical metamaterials associated with stiffness, rigidity and compressibility: A brief review. Prog. Mater. Sci. 94, 114–173. https://doi.org/10.1016/j.pmatsci.2017.12.003

Zadpoor, A.A., 2016. Mechanical meta-materials. Mater. Horiz. 3, 371–381. https://doi.org/10.1039/C6MH00065G

Zhang, P., Parnell, W.J., 2017a. Band Gap Formation and Tunability in Stretchable Serpentine Interconnects. J. Appl. Mech. 84, 091007. https://doi.org/10.1115/1.4037314

Zhang, P., Parnell, W.J., 2017b. Soft phononic crystals with deformation-independent band gaps. Proc. R. Soc. Math. Phys. Eng. Sci. 473, 20160865. https://doi.org/10.1098/rspa.2016.0865

Zingoni, A., 2009. Group-theoretic exploitations of symmetry in computational solid and structural mechanics. Int. J. Numer. Methods Eng. 79, 253–289. https://doi.org/10.1002/nme.2576

# Tables

Table 1. Point group of the high symmetry **k**-points for plane group $p4g$

|   | location | point group | operations |
|---|---|---|---|
| $\Gamma$ | $(0,0)$ | $4mm$ | $E$, $C_4^+$, $C_2$, $C_4^-$, $\sigma_x$, $\sigma_y$, $\sigma_d$, $\sigma_d'$ |
| X | $(\frac{1}{2},0)$ | $mm2$ | $E$, $C_2$, $\sigma_x$, $\sigma_y$ |
| M | $(\frac{1}{2},\frac{1}{2})$ | $4mm$ | $E$, $C_4^+$, $C_2$, $C_4^-$, $\sigma_x$, $\sigma_y$, $\sigma_d$, $\sigma_d'$ |
| $\Delta$ | $(\alpha,0)$ | $m$ | $E$, $\sigma_y$ |
| $\Sigma$ | $(\alpha,\alpha)$ | $m$ | $E$, $\sigma_d'$ |
| Y | $(\frac{1}{2},\alpha)$ | $m$ | $E$, $\sigma_x$ |

$0 < \alpha < 1/2$

Table 2. Small representation $D_{ji}^{\mathbf{k}}$ at the $\Gamma$ point for plane group $p4g$

| Label | $\{E\|\mathbf{0}\}$ | $\{C_2\|\mathbf{0}\}$ | $\{C_4^+\|\mathbf{0}\}$ | $\{C_4^-\|\mathbf{0}\}$ | $\{\sigma_x\|\boldsymbol{\tau}\}$ | $\{\sigma_y\|\boldsymbol{\tau}\}$ | $\{\sigma_d\|\boldsymbol{\tau}\}$ | $\{\sigma_d'\|\boldsymbol{\tau}\}$ |
|---|---|---|---|---|---|---|---|---|
| $\Gamma_1$ | 1 | 1 | 1 | 1 | 1 | 1 | 1 | 1 |
| $\Gamma_2$ | 1 | 1 | 1 | 1 | $-1$ | $-1$ | $-1$ | $-1$ |
| $\Gamma_3$ | 1 | 1 | $-1$ | $-1$ | 1 | 1 | $-1$ | $-1$ |
| $\Gamma_4$ | 1 | 1 | $-1$ | $-1$ | $-1$ | $-1$ | 1 | 1 |
| $\Gamma_5$ | 2 | $-2$ | 0 | 0 | 0 | 0 | 0 | 0 |
|  | **I** | $-$**I** | $i$**A** | $-i$**A** | $-$**B** | **B** | $i$**C** | $-i$**C** |

Table 3. Small representation $D_{ji}^{\mathbf{k}}$ at the $\Delta$ point for plane group $p4g$

| Label | $\{E\|\mathbf{0}\}$ | $\{\sigma_y\|\boldsymbol{\tau}\}$ |
|---|---|---|
| $\Delta_1$ | 1 | $\exp(-\alpha\pi i)$ |
| $\Delta_2$ | 1 | $-\exp(-\alpha\pi i)$ |

Table 4. Small representation $D^{\mathbf{k}}_{ji}$ at the $\Sigma$ point for plane group $p4g$

| Label | $\{E\,|\,\mathbf{0}\}$ | $\{\sigma'_d\,|\,\boldsymbol{\tau}\}$ |
|---|---|---|
| $\Sigma_1$ | 1 | $\exp(-2\alpha\pi i)$ |
| $\Sigma_2$ | 1 | $-\exp(-2\alpha\pi i)$ |

Table 5. Small representations $D^{\mathbf{k}}_{ji}$ at the $X$ point. Note that only the representative elements are given for each coset in Eq. (21).

| Label | $\{E\,|\,\mathbf{0}\}$ | $\{E\,|\,\mathbf{a}_1\}$ | $\{\sigma_y\,|\,\boldsymbol{\tau}\}$ | $\{\sigma_y\,|\,\boldsymbol{\tau}+\mathbf{a}_1\}$ | $\{C_2\,|\,\mathbf{0}\}$ | $\{C_2\,|\,\mathbf{a}_1\}$ | $\{\sigma_x\,|\,\boldsymbol{\tau}\}$ | $\{\sigma_x\,|\,\boldsymbol{\tau}+\mathbf{a}_1\}$ |
|---|---|---|---|---|---|---|---|---|
| $X_1$ | 1 | 1 | 1 | 1 | 1 | 1 | 1 | 1 |
| $X_2$ | 1 | 1 | 1 | 1 | $-1$ | $-1$ | $-1$ | $-1$ |
| $X_3$ | 1 | 1 | $-1$ | $-1$ | 1 | 1 | $-1$ | $-1$ |
| $X_4$ | 1 | 1 | $-1$ | $-1$ | $-1$ | $-1$ | 1 | 1 |
| $X_5$ | 2 | $-2$ | 0 | 0 | 0 | 0 | 0 | 0 |
|  | $\mathbf{I}$ | $-\mathbf{I}$ | $i\mathbf{A}$ | $-i\mathbf{A}$ | $-\mathbf{B}$ | $\mathbf{B}$ | $i\mathbf{C}$ | $-i\mathbf{C}$ |

Table 6. Small representations $D^{\mathbf{k}}_{ji}$ at $\mathbf{k}=(\tfrac{1}{2},\tfrac{1}{4})$. Note that only the representative elements are given for each coset in Eq. (25).

| Label | $\{E\,|\,\mathbf{0}\}$ | $\{\sigma_x\,|\,\boldsymbol{\tau}\}$ | $\{E\,|\,\mathbf{a}_2\}$ | $\{\sigma_x\,|\,\boldsymbol{\tau}+\mathbf{a}_2\}$ | $\{E\,|\,\mathbf{a}_1\}$ | $\{\sigma_x\,|\,\boldsymbol{\tau}+\mathbf{a}_1\}$ | $\{E\,|-\mathbf{a}_2\}$ | $\{\sigma_x\,|\,\boldsymbol{\tau}-\mathbf{a}_2\}$ |
|---|---|---|---|---|---|---|---|---|
| $Y_1$ | 1 | 1 | 1 | 1 | 1 | 1 | 1 | 1 |
| $Y_2$ | 1 | $\theta$ | $i$ | $-\theta^*$ | $-1$ | $-\theta$ | $-i$ | $\theta^*$ |
| $Y_3$ | 1 | $i$ | $-1$ | $-i$ | 1 | $i$ | $-1$ | $-i$ |
| $Y_4$ | 1 | $-\theta^*$ | $-i$ | $\theta$ | $-1$ | $\theta^*$ | $i$ | $-\theta$ |
| $Y_5$ | 1 | $-1$ | 1 | $-1$ | 1 | $-1$ | 1 | $-1$ |
| $Y_6$ | 1 | $-\theta$ | $i$ | $\theta^*$ | $-1$ | $\theta$ | $-i$ | $-\theta^*$ |
| $Y_7$ | 1 | $-i$ | $-1$ | $i$ | 1 | $-i$ | $-1$ | $i$ |
| $Y_8$ | 1 | $\theta^*$ | $-i$ | $-\theta$ | $-1$ | $-\theta^*$ | $i$ | $\theta$ |

$\theta=\exp(i\pi/4),\ \theta^*=\exp(-i\pi/4)$

Table 7. Small representations $D_{ji}^{\mathbf{k}}$ at the $M$ point. Note that only the representative elements are given for each coset in Eq. (28).

| Label | $\{E\|\mathbf{0}\}$ | $\{C_2\|\mathbf{0}\}$ | $\{C_4^+\|\mathbf{0}\}$ | $\{C_4^-\|\mathbf{0}\}$ | $\{\sigma_x\|\boldsymbol{\tau}\}$ | $\{\sigma_y\|\boldsymbol{\tau}\}$ | $\{\sigma_d\|\boldsymbol{\tau}\}$ | $\{\sigma_d'\|\boldsymbol{\tau}\}$ | $\{E\|\mathbf{a}_1\}$ | $\{C_2\|\mathbf{a}_1\}$ | $\{C_4^+\|\mathbf{a}_1\}$ | $\{C_4^-\|\mathbf{a}_1\}$ | $\{\sigma_x\|\boldsymbol{\tau}+\mathbf{a}_1\}$ | $\{\sigma_y\|\boldsymbol{\tau}+\mathbf{a}_1\}$ | $\{\sigma_d\|\boldsymbol{\tau}+\mathbf{a}_1\}$ | $\{\sigma_d'\|\boldsymbol{\tau}+\mathbf{a}_1\}$ |
|---|---|---|---|---|---|---|---|---|---|---|---|---|---|---|---|---|
| $M_1$ | 1 | 1 | 1 | 1 | 1 | 1 | 1 | 1 | 1 | 1 | 1 | 1 | 1 | 1 | 1 | 1 |
| $M_2$ | 1 | 1 | −1 | −1 | 1 | 1 | −1 | −1 | 1 | 1 | −1 | −1 | 1 | 1 | −1 | −1 |
| $M_3$ | 1 | 1 | −1 | −1 | −1 | −1 | 1 | 1 | 1 | 1 | −1 | −1 | −1 | −1 | 1 | 1 |
| $M_4$ | 1 | 1 | 1 | 1 | −1 | −1 | −1 | −1 | 1 | 1 | 1 | 1 | −1 | −1 | −1 | −1 |
| $M_5$ | 1 | −1 | $i$ | $-i$ | $i$ | $-i$ | 1 | −1 | −1 | 1 | $-i$ | $i$ | $-i$ | $i$ | −1 | 1 |
| $M_6$ | 1 | −1 | $-i$ | $i$ | $-i$ | $i$ | 1 | −1 | −1 | 1 | $i$ | $-i$ | $i$ | $-i$ | −1 | 1 |
| $M_7$ | 1 | −1 | $-i$ | $i$ | $i$ | $-i$ | −1 | 1 | −1 | 1 | $i$ | $-i$ | $-i$ | $i$ | 1 | −1 |
| $M_8$ | 1 | −1 | $i$ | $-i$ | $-i$ | $i$ | −1 | 1 | −1 | 1 | $-i$ | $i$ | $i$ | $-i$ | 1 | −1 |
| $M_9$ | 2 2 $\mathbf{I}$ $\mathbf{I}$ | 2 2 $\mathbf{I}$ $\mathbf{I}$ | 0 0 $i\mathbf{C}$ $i\mathbf{C}$ | 0 0 $i\mathbf{C}$ $i\mathbf{C}$ | 0 0 $i\mathbf{A}$ $i\mathbf{A}$ | 0 0 $i\mathbf{A}$ $i\mathbf{A}$ | 0 0 $-\mathbf{B}$ $-\mathbf{B}$ | 0 0 $-\mathbf{B}$ $-\mathbf{B}$ | −2 −2 $-\mathbf{I}$ $-\mathbf{I}$ | −2 −2 $-\mathbf{I}$ $-\mathbf{I}$ | 0 0 $-i\mathbf{C}$ $-i\mathbf{C}$ | 0 0 $-i\mathbf{C}$ $-i\mathbf{C}$ | 0 0 $-i\mathbf{A}$ $-i\mathbf{A}$ | 0 0 $-i\mathbf{A}$ $-i\mathbf{A}$ | 0 0 $\mathbf{B}$ $\mathbf{B}$ | 0 0 $\mathbf{B}$ $\mathbf{B}$ |
| $M_{10}$ | 2 2 $\mathbf{I}$ $\mathbf{I}$ | −2 −2 $-\mathbf{I}$ $-\mathbf{I}$ | 0 0 $i\mathbf{B}$ $i\mathbf{B}$ | 0 0 $-i\mathbf{B}$ $-i\mathbf{B}$ | 0 0 $i\mathbf{C}$ $i\mathbf{C}$ | 0 0 $-i\mathbf{C}$ $-i\mathbf{C}$ | 0 0 $\mathbf{A}$ $\mathbf{A}$ | 0 0 $-\mathbf{A}$ $-\mathbf{A}$ | 2 2 $\mathbf{I}$ $\mathbf{I}$ | −2 −2 $-\mathbf{I}$ $-\mathbf{I}$ | 0 0 $i\mathbf{B}$ $i\mathbf{B}$ | 0 0 $-i\mathbf{B}$ $-i\mathbf{B}$ | 0 0 $i\mathbf{C}$ $i\mathbf{C}$ | 0 0 $-i\mathbf{C}$ $-i\mathbf{C}$ | 0 0 $\mathbf{A}$ $\mathbf{A}$ | 0 0 $-\mathbf{A}$ $-\mathbf{A}$ |

Table 8. Small representation $D_{ji}$ at the $\Delta$ point when the periodic Bloch function is used.

| Label | $\{E\|\mathbf{0}\}$ | $\{\sigma_y\|\boldsymbol{\tau}\}$ |
|---|---|---|
| $\Delta_1$ | 1 | 1 |
| $\Delta_2$ | 1 | −1 |

Table 9. Small representation $D_{ji}$ at the $\Sigma$ point when the periodic Bloch function is used.

| Label | $\{E\|\mathbf{0}\}$ | $\{\sigma_d'\|\boldsymbol{\tau}\}$ |
|---|---|---|
| $\Sigma_1$ | 1 | 1 |
| $\Sigma_2$ | 1 | −1 |

# Figure Captions

Figure 1. Schematic illustrations of some representative structures with glide symmetry. (a) Hinged rotating squares. (b) Porous sheet. (c) Sinusoidal lattice. (d) Anti-tetrachiral structure. (e) Undulating structure. (f) 'Anti-rolls' buckling mode of honeycomb structures. The four structures in (a)-(d) belong to the plane group *p*4*g*, while the structures in (e) and (f) are in the Frieze group *p*2*mg* and plane group *pgg*, respectively. The graphs in (c) and (f) are reproduced from (Haghpanah et al., 2014).

Figure 2. Sinusoidal lattice structure with glide symmetry. Both primitive and non-primitive unit cells are sketched by using dotted lines. In this work, we adopt the primitive unit cell for the analysis.

Figure 3. Unit cells in the real and **k**-space. (a) Symmetry operations of the lattice structure in the real space. A total of four types of glide symmetry exist for this structure. (b) The FBZ and irreducible FBZ (shaded area) in the **k**-space. The reciprocal lattice vectors are indicated as $\mathbf{b}_1$ and $\mathbf{b}_2$, respectively.

Figure 4. Phononic band structure of the *p*4*g* lattice structure in the irreducible FBZ. The type of small representation is labelled for each band to show the symmetry of the corresponding phonon modes. Each band is depicted in a different color. Double degeneracies can be found at the center and boundary of the FBZ.

Figure 5. Phonon modes with different representations at $\Gamma$ point. The undeformed lattice structure is indicated by dashed curves while the phonon modes are depicted by solid curves. The phonon modes are periodic for each unit cell at the $\Gamma$ point.

Figure 6. Phonon modes with different representations at the $\Delta$ and $\Sigma$ points.

Figure 7. Phonon modes with different representations at the *X* and *Y* points.

Figure 8. Phonon modes with different representations at the *M* point.

Figure 9. Periodic Bloch functions of the phonon modes at the $\Delta$ and $\Sigma$ points.

Figure 1

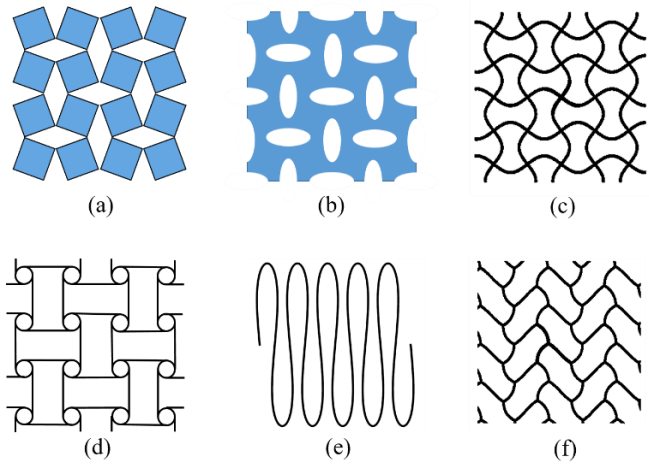

Figure 2

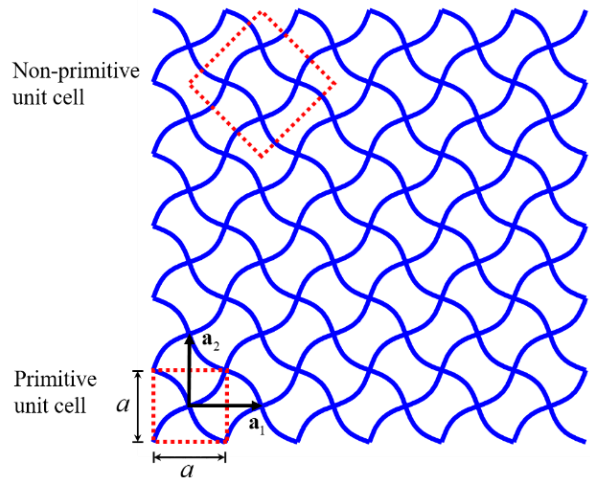

Figure 3

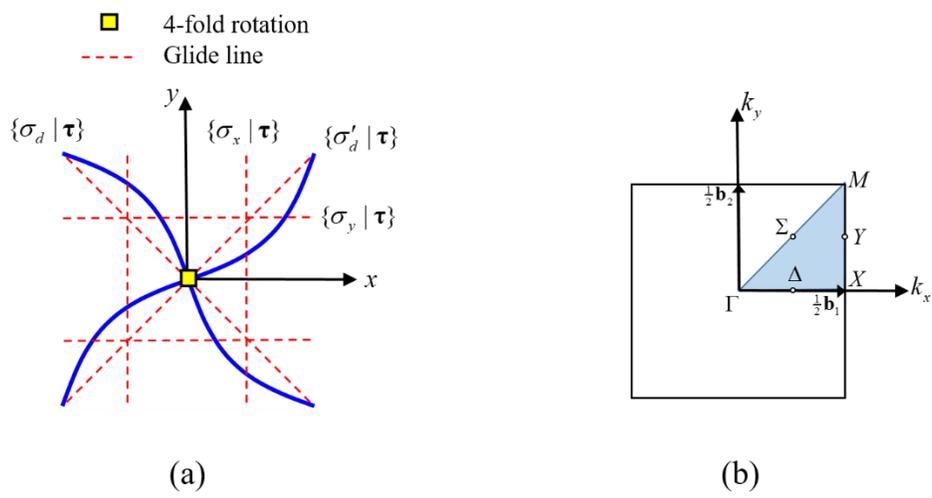

(a)                          (b)

Figure 4

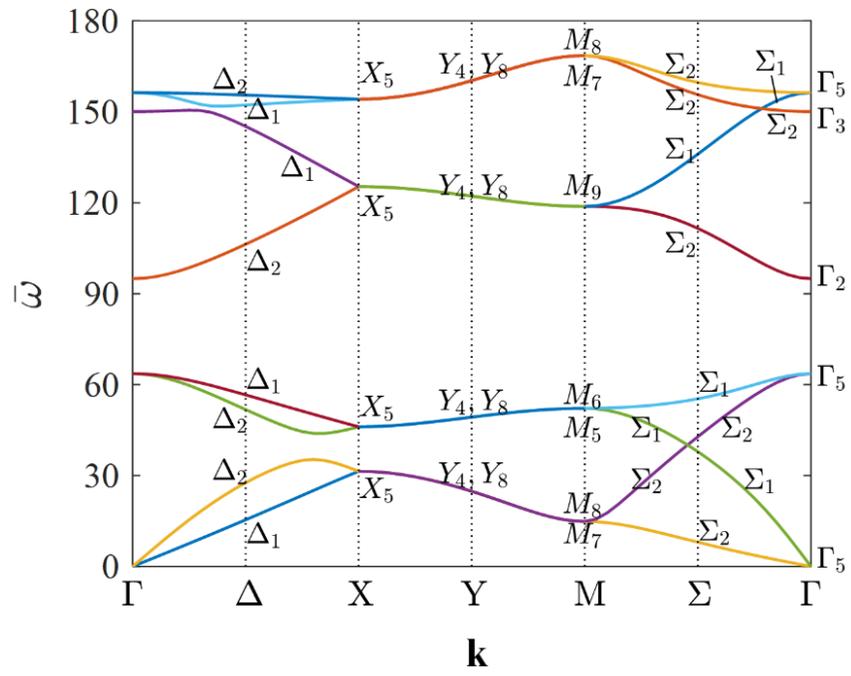

Figure 5

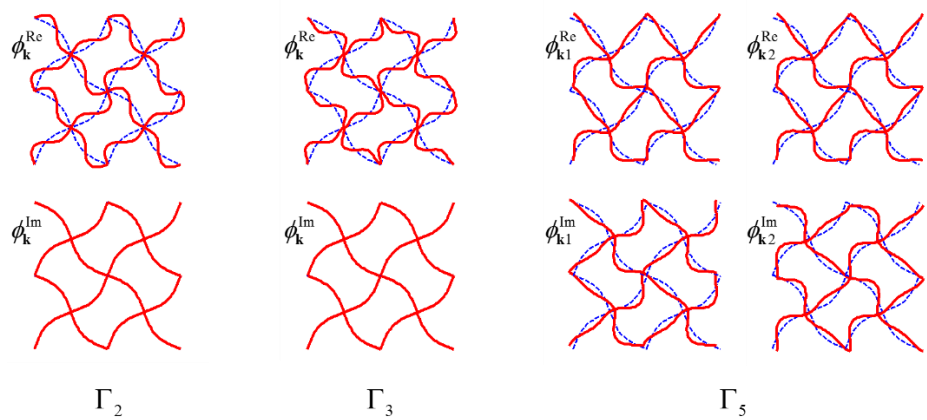

$\Gamma_2$ $\quad\quad\quad\quad\quad$ $\Gamma_3$ $\quad\quad\quad\quad\quad$ $\Gamma_5$

Figure 6

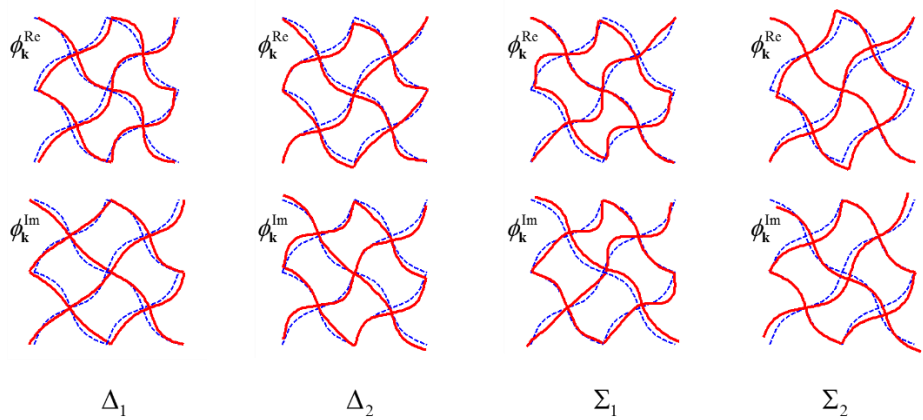

Figure 7

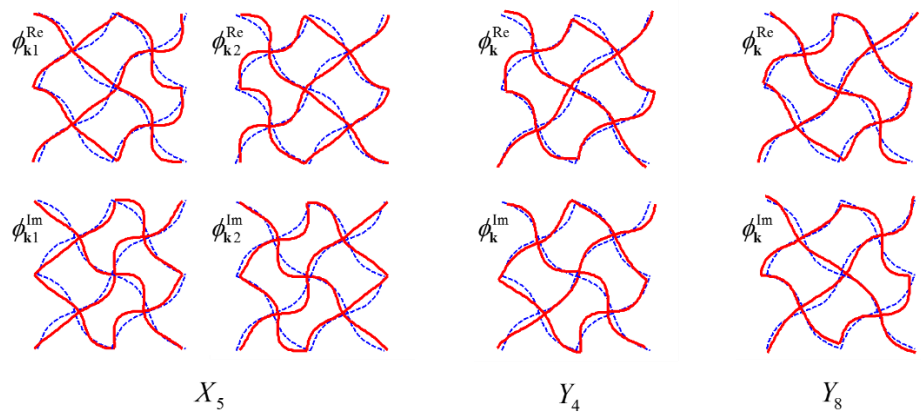

$X_5$        $Y_4$        $Y_8$

Figure 8

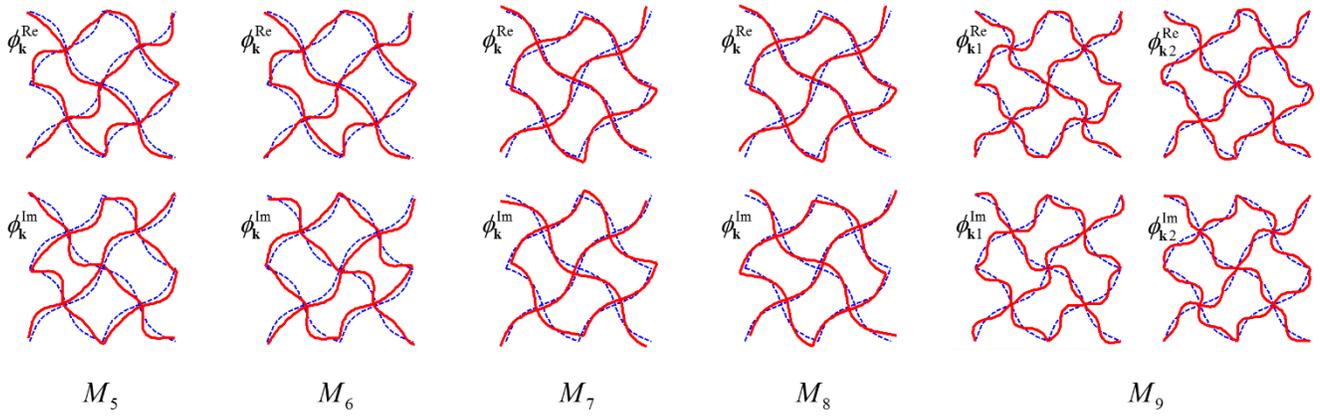
$M_5 \qquad M_6 \qquad M_7 \qquad M_8 \qquad M_9$

Figure 8

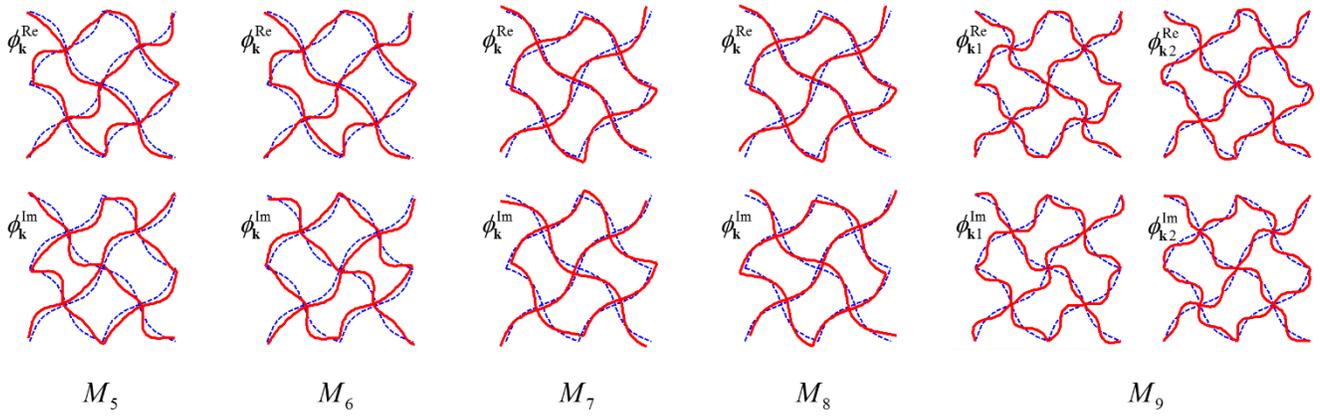

$M_5 \qquad M_6 \qquad M_7 \qquad M_8 \qquad M_9$

Figure 9

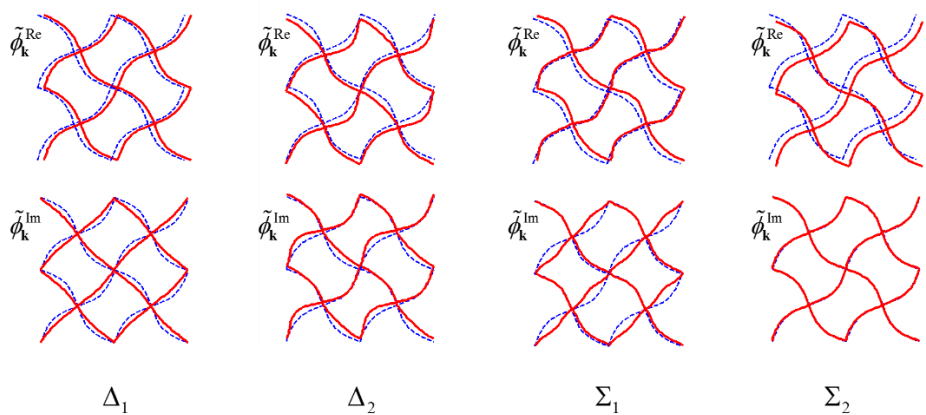